\documentclass[12pt]{article}

\usepackage{amssymb,esvect,amsmath,graphicx,latexsym,amsthm,slashed,eso-pic,hyperref}

\usepackage{epsfig,amsmath,amssymb,verbatim,mathrsfs,hyperref}
\usepackage{xspace}
\usepackage{xcolor}
\usepackage{slashed}
\usepackage{dcolumn}
\usepackage{multirow}
\usepackage{color}
\graphicspath{{./}{./Figs/}{./Figures/}}

\def\beq{\begin{eqnarray}}
\def\eeq{\end{eqnarray}}
\def\bea{\begin{eqnarray}}
\def\eea{\end{eqnarray}}

\def\tev{\, {\rm TeV}}
\def\gev{\, {\rm GeV}}
\def\mev{\, {\rm MeV}}

\newcommand{\gsim}{\lower.7ex\hbox{$\;\stackrel{\textstyle>}{\sim}\;$}}
\newcommand{\lsim}{\lower.7ex\hbox{$\;\stackrel{\textstyle<}{\sim}\;$}}

\newcommand{\ccdot}{\!\cdot\!}
\newcommand{\nnmb}{\nonumber}
\newcommand{\del}{\partial}
\newcommand{\met}{\,/\hspace{-0.25cm}E_T}

\newcommand{\lrf}[2]{\left(\frac{#1}{#2}\right)}

\newcommand{\lag}{\mathscr{L}}

\newcommand{\BR}{\mathrm{BR}}
\newcommand{\fsl}[1]{\backslash\hspace{-0.2cm}{#1}}

\begin{document}
\begin{titlepage}
\noindent
\begin{center}
  \begin{Large}
    \begin{bf}
Higgs Boson Decays to Dark Photons\vspace{0.15cm}\\
through the\vspace{0.37cm}\\ 
Vectorized Lepton Portal
     \end{bf}
  \end{Large}
\end{center}
\vspace{0.2cm}
\begin{center}
\begin{large}
Qianshu~Lu$^{(a,b)}$, David~E.~Morrissey$^{(a)}$, and Alexander~M.~Wijangco$^{(a)}$
\end{large}
\vspace{1cm}\\
\begin{it}
(a) TRIUMF, 4004 Wesbrook Mall, Vancouver, BC, Canada V6T 2A3
\vspace{0.2cm}\\
(b) Division of Engineering Science, University of Toronto, Toronto, ON, Canada M5S 2E4
\vspace{0.5cm}\\
email: 
\emph{\texttt{qianshu.lu@mail.utoronto.ca}},
\emph{\texttt{dmorri@triumf.ca}},
\emph{\texttt{awijangco@triumf.ca}}
\vspace{0.2cm}
\end{it}
\end{center}
\center{\today}

\begin{abstract}

  Vector-like fermions charged under both the Standard Model and
a new dark gauge group arise in many theories of new physics.  If these
fermions include an electroweak doublet and singlet with equal dark
charges, they can potentially connect to the Higgs field through 
a Yukawa coupling in analogy to the standard neutrino portal.  
With such a coupling, fermion loops generate exotic decays 
of the Higgs boson to one or more dark vector bosons.
In this work we study a minimal realization of this scenario with 
an Abelian dark group.  We investigate the potential new Higgs decays modes,
we compute their rates, and we study the prospects for observing
them at the Large Hadron Collider and beyond given the other
experimental constraints on the theory.  We also discuss extensions
of the theory to non-Abelian dark groups.

\end{abstract}

\end{titlepage}

\setcounter{page}{2}


\section{Introduction\label{sec:intro}}

  Dark sectors have been studied extensively in recent 
years~\cite{Fayet:2007ua,Pospelov:2008zw,Bjorken:2009mm,Essig:2013lka,Alexander:2016aln}.
Such sectors consist of new states that interact only very weakly 
with the Standard Model~(SM).  This allows the new physics in the dark 
sector to be relatively light, with characteristic mass well below 
the electroweak scale, while still being consistent with current 
experimental tests.  Dark sectors may also be related
to (or comprise) the dark matter in the universe~\cite{Boehm:2003hm,Borodatchenkova:2005ct,Pospelov:2007mp,ArkaniHamed:2008qn}.

  While the range of possibilities for dark sectors is enormous,
particular attention has been given to those that connect
to the SM through a set of portal operators:
\beq
\text{Vector Portal}:&& \frac{\epsilon}{2}B_{\mu\nu}X^{\mu\nu}
\label{eq:vport}\\
\text{Higgs Portal}:&& (A\phi + \kappa\phi^2)|H|^2\phantom{\frac{1}{2}}
\label{eq:hport}
\\
\text{Neutrino Portal}:&& y_N\bar{L}\widetilde HN\phantom{\frac{1}{2}}
\label{eq:nport}
\eeq  
First, in the vector portal, a new Abelian vector boson $X$ couples
to the SM through kinetic mixing with 
hypercharge~\cite{Okun:1982xi,Holdom:1985ag}.
Second, in the Higgs portal~\cite{Schabinger:2005ei,Patt:2006fw}, 
a new scalar connects with the SM Higgs field.
And third, in the neutrino portal a new gauge singlet fermion $N$
connects to the SM lepton and Higgs doublets.
These portals represent the three ways in which a new field with no
SM charges can couple to the SM at the renormalizable level. 
As such, these interactions are non-decoupling,
and the most sensitive searches for light new physics connecting to us
through these interactions are typically lower-energy experiments
with very high intensity or precision~\cite{Essig:2013lka}.

  The portal interactions of Eqs.~(\ref{eq:vport},\ref{eq:hport},\ref{eq:nport})
can be generated by integrating out massive \emph{mediator} states 
that couple directly to both the visible and dark sectors.  Such mediators
can give rise to new and unusual signals at high-energy colliders such
as the Large Hadron Collider~(LHC), either through their direct production 
or by providing a new avenue to populate the light states in the dark 
sector~\cite{Strassler:2006im,Strassler:2006qa,Han:2007ae}.  
Discovering mediator particles or measuring their decoupling effects
would also provide new insight into the structure and dynamics of the light states
in the dark sector.

  In this paper we investigate a very simple theory of mediators to a 
dark sector consisting of a $U(1)_x$ vector boson $X$, first
presented in Ref.~\cite{Davoudiasl:2012ig}.
The mediators are an electroweak singlet $N$ and doublet $P$ 
of Dirac fermions with hypercharges $Y=0,\,-1/2$ and equal dark charge $q_x$. 
These quantum numbers allow for vector-like fermion masses
and a coupling to the SM Higgs boson of the form
\beq
-\lag ~\supset~ \lambda\,\overline{P}\widetilde{H}N + (h.c.) \ ,
\label{eq:vlport}
\eeq
where $\widetilde{H}= i\sigma_2H^*$.  After electroweak symmetry breaking,
the singlet and doublet mix to form a pair of neutral Dirac fermions 
$\psi_1$ and $\psi_2$, and a charged fermion $P^-$.
We assume as well that the dark vector boson develops a mass $m_x$, 
either through a dark sector Higgs or Stueckelberg 
mechanism~\cite{Stueckelberg:1900zz,Kors:2004dx}.
The interaction of Eq.~\eqref{eq:vlport} is analogous to the neutrino portal, 
but it involves the new $U(1)_x$ charged mediators instead of the SM leptons;
we call it the \emph{vectorized lepton portal}.  
In addition to this portal interaction, loops of the new fermions
also contribute to a vector portal coupling between the $U(1)_x$ vector boson 
and hypercharge.  

This general structure appears in a broad range of proposed
extensions of the SM.
The new fermions in the theory have the same SM quantum numbers
as some of the models of vector-like leptons (without dark charges)
considered in Refs.~\cite{Ellis:2014dza,Bhattacharya:2015qpa}.
In theories with supersymmetry, superpotential couplings of the form
of Eq.~\eqref{eq:vlport} are the origin of general renormalizable 
Higgs portal interactions via scalar $F$-terms, and they have been
invoked to connect the Higgs to gauge mediator 
supermultiplets~\cite{Dvali:1996cu,Craig:2012xp} and to  increase 
the mass of the SM-like Higgs 
boson~\cite{Azatov:2011ht,Heckman:2011bb,Evans:2012uf}.
Closely related structures with non-Abelian dark gauge groups
also emerge in some theories of neutral naturalness
such as folded supersymmetry~\cite{Burdman:2006tz} and quirky 
little Higgs~\cite{Cai:2008au}, and in relaxion 
constructions~\cite{Graham:2015cka,Antipin:2015jia,Batell:2015fma,Choi:2015fiu}.
Realizations of the vectorized lepton portal with an Abelian dark group
were studied in Refs.~\cite{Davoudiasl:2012ig,Davoudiasl:2013aya,DiFranzo:2015nli,DiFranzo:2016uzc}, and with a non-Abelian group in Ref.~\cite{Juknevich:2009gg,Beauchesne:2017ukw}.

  The vectorized lepton portal can induce a wide range of new experimental
signals, both from the light dark vector and the heavier mediator fermions.  
The new signals of primary interest in this work are exotic decays
of the SM Higgs boson.  Loops of the vector-like fermions give rise
to $h\to XX$ and $h\to XZ$ decay channels.  We show that the resulting
branching fractions can be much larger than from kinetic 
mixing alone.  Furthermore, we also demonstrate that these decays are potentially 
observable at the LHC (and beyond) while being consistent with current
bounds from precision electroweak tests and direct collider searches.
Relative to the closely related previous works of 
Refs.~\cite{Davoudiasl:2012ig,Davoudiasl:2013aya}, we compute the Higgs
decay widths and the direct constraints due to the new fermions in more detail,
and we show that current direct limits allow for observable Higgs signals
at the LHC.

  Following this introduction, we present a simple vectorized
lepton portal model in more detail in Sec.~\ref{sec:setup}.
Next, we calculate the Higgs boson decay widths to dark vectors through mediator
fermion loops and discuss their observability at the LHC and beyond
in Sec.~\ref{sec:higgs}.  Constraints on the mediators
from precision electroweak measurements, direct searches at the LHC,
and stability of the Higgs potential are discussed in Sec.~\ref{sec:bounds}.
In Sec.~\ref{sec:dm} we study the implications
of the theory for dark matter and cosmology, and we discuss some
potential extensions of the minimal theory motivated by them.
Further extensions of the minimal theory to non-Abelian dark gauge
groups are discussed in Sec.~\ref{sec:nonab}.
Finally, we reserve Sec.~\ref{sec:conc} for our conclusions.

\section{Fields, Masses, and Interactions\label{sec:setup}}

   We consider a theory with two new vector-like fermion multiplets 
with charge assignments under $SU(3)_c\times SU(2)_L\times U(1)_Y\times U(1)_x$
of $N = (1,1,0;\,q_x)$ and $P=(1,2,-1/2;\,q_x)$. 
This allows the Yukawa coupling and masses:
\beq
-\lag \supset \left(\lambda\overline{P}\widetilde{H}N + h.c.\right)
+ m_P\overline{P}P + m_N\overline{N}N \ ,
\label{eq:vlep1}
\eeq
where $\widetilde{H} = i\sigma_2H^*$.  Note that $m_P$, $m_N$, and $\lambda$ 
can all be taken to be real and positive through field redefinitions.
We also normalize the dark gauge coupling $g_x$ such that 
either $q_x= 1$ or $q_x = -1$.  The set of fermion charges in our theory
is minimal in that there is only one new (Dirac) field with SM gauge
charges.  Let us also mention that the Yukawa interaction of Eq.~\eqref{eq:vlep1}
can be generalized to a chiral form with two independent Yukawa 
couplings that allows for $CP$ 
violation~\cite{Mahbubani:2005pt,DEramo:2007anh,Essig:2007az,Cohen:2011ec};
we focus on the parity-preserving form of Eq.~\eqref{eq:vlep1} for simplicity.

\subsection{Minimal Masses and Interactions}

Expanding the Higgs about its vacuum expectation value~(VEV) in unitary gauge, 
$H \to (v+h/\sqrt{2})$, with $v=174\,\gev$, and writing the $SU(2)_L$ components 
of the doublet explicitly as $P = (P^0,\,P^-)^T$, the fermion terms become
\beq
-\lag \supset -m_P\overline{P}^-P^-
+ \left(\overline{N},\overline{P}^0\right)\!\left(\begin{array}{cc}
m_N&\lambda\,v\\
\lambda\,v&m_P
\end{array}\right)\!
\left(\begin{array}{c}N\\P^0\end{array}\right)
+ \frac{\lambda}{\sqrt{2}}h\,\left(\overline{N}P^0 +\overline{P}^0N\right) \ .
\eeq
The physical states are therefore a charged fermion $P^-$ 
with mass $m_P$ together with two SM-neutral Dirac fermions 
$\psi_{1,2}$ with masses
\beq
m_{1,2} = \frac{1}{2}\left[(m_N+m_P) \mp \sqrt{(m_N-m_P)^2+4\lambda^2v^2}\right]
 \ .
\label{eq:mass}
\eeq
We only consider solutions with positive $m_1 > 0$ in this work,
corresponding to the condition $\sqrt{m_Nm_P} > \lambda v$, since the $m_1<0$
solution has $|m_1| \leq \lambda\,v$ and is strongly 
constrained by direct searches.
The neutral gauge eigenstates are related to the mass eigenstates by
\beq
\left(
\begin{array}{c}
N\\P^{0}
\end{array}
\right)
= 
\left(
\begin{array}{cc}
c_{\alpha}&s_{\alpha}\\
-s_{\alpha}&c_{\alpha}
\end{array}
\right)
\left(
\begin{array}{c}
\psi_1\\\psi_2
\end{array}
\right) \ .\label{eq:mix1}
\eeq
with the mixing angle given by
\beq
\tan(2\alpha) = \frac{2\lambda {v}}{m_P-m_N} \ .
\label{eq:mix2}
\eeq
We choose the solution for $\alpha$ such that $m_1< m_2$.  

  Rewriting the Yukawa interaction in terms of the mass eigenstates,
we find
\beq
-\lag ~\supset~ \frac{\lambda}{\sqrt{2}}\,h\,\left[
2s_{\alpha}c_{\alpha}\,
(-\overline{\psi}_1\psi_1 +\overline{\psi}_2\psi_2 )
+ (c_{\alpha}^2-s_{\alpha}^2)\,(\overline{\psi}_1\psi_2 +\overline{\psi}_2\psi_1)
\right] \ .
\eeq
Note that the charged $P^-$ state does not couple to the Higgs boson.
The relevant vector boson couplings are
\beq
-\lag  &\supset&
\bar{g}(-\frac{1}{2}+s_W^2)Z_{\mu}\,\overline{P}^-\!\gamma^{\mu}P^-
- eA_{\mu}\,\overline{P}^-\!\gamma^{\mu}P^-\nnmb\\
&&
+\frac{g}{\sqrt{2}}\left[W_{\mu}^+\overline{P}^-\!\gamma^{\mu}(-s_{\alpha}\psi_1
+c_{\alpha}\psi_2) + (h.c.)\right]\label{eq:dcurrent2c1}\\
&&
+ \frac{1}{2}\bar{g}Z_{\mu}\left[
s_{\alpha}^2\overline{\psi}_1\gamma^{\mu}\psi_1
+c_{\alpha}^2\overline{\psi}_2\gamma^{\mu}\psi_2
-s_{\alpha}c_{\alpha}(\overline{\psi}_1\gamma^\mu\psi_2
+\overline{\psi}_2\gamma^\mu\psi_1)\right]\nnmb\\
&&
+ g_xX_{\mu}\left[\overline{\psi}_1\gamma_{\mu}\psi_1
+\overline{\psi}_2\gamma_{\mu}\psi_2
+\overline{P}^-\!\gamma_{\mu}P^-\right]
\ , \nnmb
\eeq
where $\bar{g} = g/c_W = \sqrt{g^2+{g^\prime}^2}$.

  Beyond the Yukawa and gauge couplings above, the dark sector also
couples to the SM through gauge kinetic mixing~\cite{Okun:1982xi,Holdom:1985ag},
\beq
-\lag ~\supset~ \frac{\epsilon}{2c_W}B_{\mu\nu}X^{\mu\nu} \ .
\eeq
This interaction can be treated as 
in Refs.~\cite{Pospelov:2008zw,Bjorken:2009mm,Davoudiasl:2012ag},
with the main effect for $m_x \ll m_Z$ being kinetic mixing with the photon
with strength $\epsilon$.  It allows the dark vector to decay to lighter
SM final states.

We take $\epsilon$ to be an independent
parameter, but it should be noted that it is generated by $P$ loops.  
The log-enhanced running contribution to $\epsilon$
from these loops between scale $\mu$ and $m_P$ 
is~\cite{delAguila:1988jz,Dienes:1996zr}
\beq
\Delta\epsilon &\simeq&  
-\frac{q_x}{3\pi}\sqrt{\alpha_x\alpha}\,\ln\lrf{\mu}{m_P} 
\\
&\simeq& -q_x\,(3\times 10^{-3})\,\lrf{\alpha_x}{10\alpha}^{1/2}\,
\ln\!\lrf{\mu}{m_P} \ ,
\nnmb
\eeq
where $\alpha_x = g_x^2/4\pi$.
Values much smaller than this are expected to require some degree of tuning,
or additional structure in the theory such as an approximately conserved
charge conjugation symmetry in the dark sector~\cite{DiFranzo:2015nli}.

\subsection{Additional Interactions}

  Several other interactions can be added to the minimal set
discussed above if the dark sector contains a scalar $\phi$ with dark charge
$Q_x$, such as a dark Higgs boson responsible for generating the 
dark vector mass~\cite{Davoudiasl:2012ig,Davoudiasl:2013aya}.  
For any $Q_x$, the scalar can connect to the SM Higgs 
field through the Higgs portal,
\beq
-\lag ~\supset~ \kappa|\phi|^2|H|^2 \ .
\eeq
This will induce Higgs mixing if $\phi$ develops a VEV.
Such an interaction is generated at two-loop order through the
gauge and Yukawa couplings of the theory with size
\beq
\Delta\kappa &\sim& \frac{Q_x^2}{(4\pi)^2}\lambda^2\alpha_x^2 
\\
&=& (4\times 10^{-5})\,\lrf{\alpha_x}{10\alpha}^2\lambda^2Q_x^2
\nnmb \ .
\eeq
As for $\epsilon$, we take this as a lower limit on the natural size 
of $\kappa$.  It is parametrically smaller than the sizes of the effects
we consider.

  Other gauge invariant operators are possible for special values 
of the charge $Q_x$ of $\phi$~\cite{Davoudiasl:2012ig,Davoudiasl:2013aya}.  
For $Q_x = \pm q_x$, a direct lepton mixing is allowed,
\beq
-\lag \supset y_L\phi\,\overline{P}_RL_L + (h.c.) \ ,
\label{eq:lcoup}
\eeq 
where $L_L$ is the SM lepton doublet.  
This operator can contribute to lepton masses and flavor violation,
but current bounds can typically be satisfied for couplings below
$|y_L|\lesssim 10^{-3}$~\cite{Davoudiasl:2012ig}.
With $Q_x=-2q_x$ we can write
\beq
-\lag &\supset& y_N\phi\,\overline{N^c}N  + (h.c.) \ ,
\label{eq:maj}
\eeq
which induces a Majorana mass for $N$ 
(and a one-loop contribution to $\kappa$)
for non-zero $\langle\phi\rangle$.

\section{Higgs Boson Decays to Dark Vectors\label{sec:higgs}}

  Decays of the Higgs boson to one or more dark vectors are generated
by the portal coupling of Eq.~\eqref{eq:vlport}.  These arise at one-loop
order from UV-finite triangle diagrams, in direct analogy to the contributions
to the SM Higgs decay modes $h\to\gamma\gamma$ and $h\to gg$ from loops of the 
top quark.  In our minimal vectorized lepton portal scenario,
the new decay channels are $h \to XX$ and $h\to XZ$.  We investigate
these decays in this section.

  Before proceeding, let us also mention that the mediator fermions typically 
do not modify the Higgs branching fractions to SM final states in 
a significant way.  There is no direct one-loop contribution to $h\to gg$ 
since the mediators are uncolored, and the absence of a tree-level coupling 
of the charged $P^-$ mode to the Higgs implies the same for 
$h\to \gamma\gamma$ and $h\to \gamma X$.  
The primary exception to this occurs when the new fermions are light enough
that $h\to \psi_1\overline{\psi}_1$ is allowed.  For $\lambda$ of order
unity, this channel can easily dominate the Higgs width.  Since our
focus is on decays of the Higgs to dark vectors, which require
larger $\lambda$ to be relevant, we concentrate on fermion masses
greater than $m_1 > m_h/2$.

\subsection{Higgs Branching Fractions}

The vectorized lepton portal can induce both $h\to XX$ and $h\to XZ$ decays 
at a similar level.
We collect in Appendix~\ref{app:hloop} the loop functions
relevant for the decay.  The asymptotic form of the $h\to XX$ decay
in the limit $m_{1,2}\gg m_h$ and $m_x\to 0$ can be obtained as a 
low-energy Higgs theorem~\cite{Shifman:1979eb}.  The result is
\beq
\lag_{eff} &\supset& -\frac{1}{4}\,\frac{\alpha_x}{4\pi}
\left(\sum_{i=1}^2\Delta b_i\frac{2}{m_i}\frac{\del m_i}{\del v}\right)\!
\lrf{h}{\sqrt{2}}X_{\mu\nu}X^{\mu\nu}\\
&=& -\frac{\alpha_x}{3\pi}\frac{\lambda^2v}{m_1m_2}
\lrf{h}{\sqrt{2}}X_{\mu\nu}X^{\mu\nu} \ ,
\eeq
where $\Delta b_i = -4/3$, 
and corresponds to the gauge invariant effective operator
\beq
\mathcal{L}_{eff} \supset 
-\frac{\alpha_x}{6\pi}\frac{\lambda^2}{m_1m_2\,}H^{\dagger}H\,
X_{\mu\nu}X^{\mu\nu} \ .
\label{eq:heft}
\eeq
We find the same result from the appropriate limit of the full 
loop calculation.\footnote{
Our result is smaller by a factor of two than the related calculation 
of Ref.~\cite{Juknevich:2009gg}.} 
The expression of Eq.~\eqref{eq:heft} shows that in the heavy fermion limit,
the $h\to XX$ decay amplitude depends quadratically on the lepton portal
Yukawa coupling $\lambda$ and the dark gauge coupling $g_x$, 
and decouples if either of the neutral modes becomes very heavy. 
Both features arise from the non-diagonal
Higgs coupling to the $P$ and $N$ fields.  A similar low-energy
calculation can be performed for the $h\to XZ$ mode, but the result 
is less illuminating and does not correspond to a single gauge-invariant operator.
However, the result scales approximately quadratically in $\lambda$ 
and linearly in $g_x$.

\begin{figure}[ttt]
 \begin{center}
         \includegraphics[width = 0.434\textwidth]{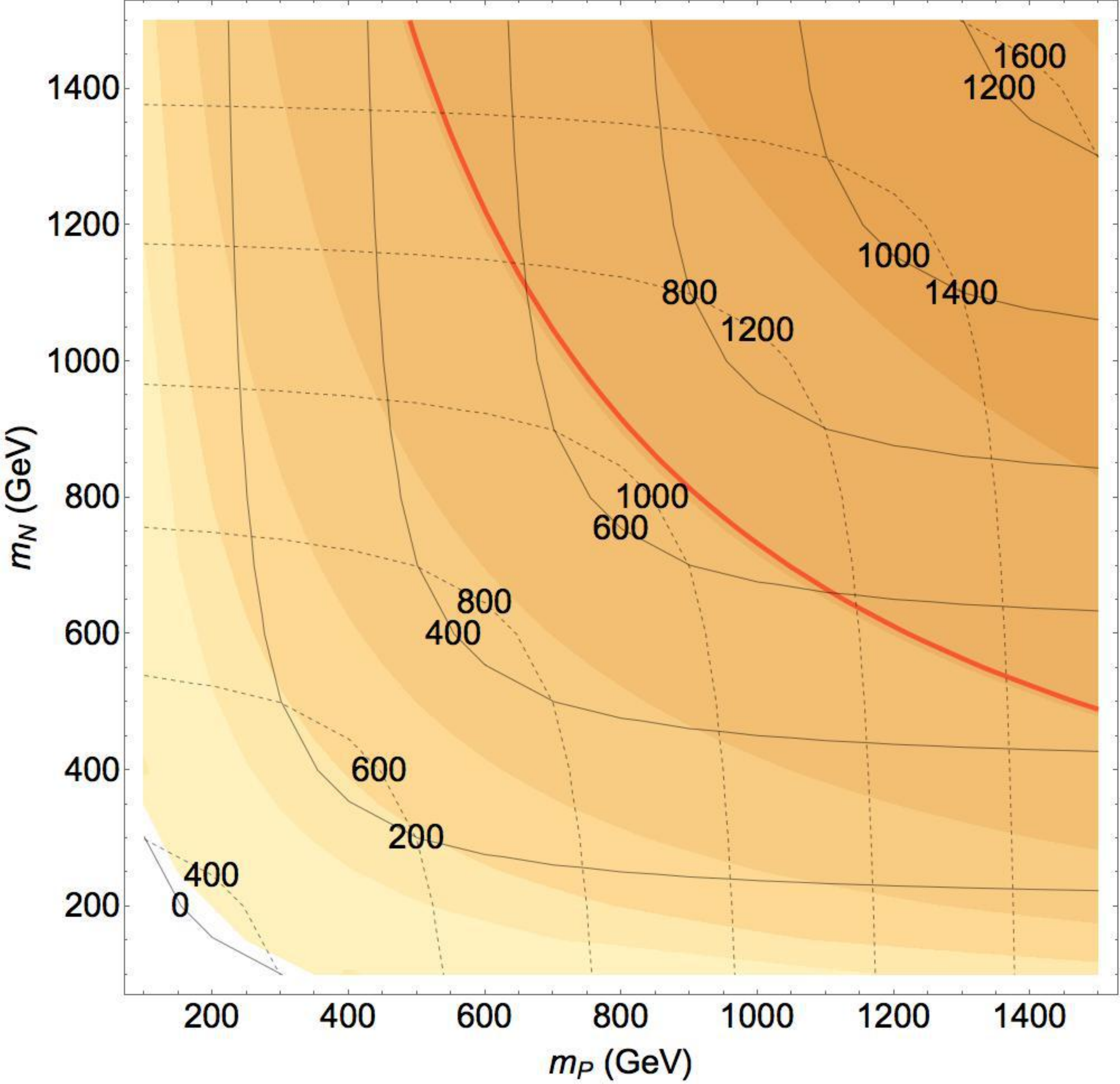}~~~
         \includegraphics[width = 0.483\textwidth]{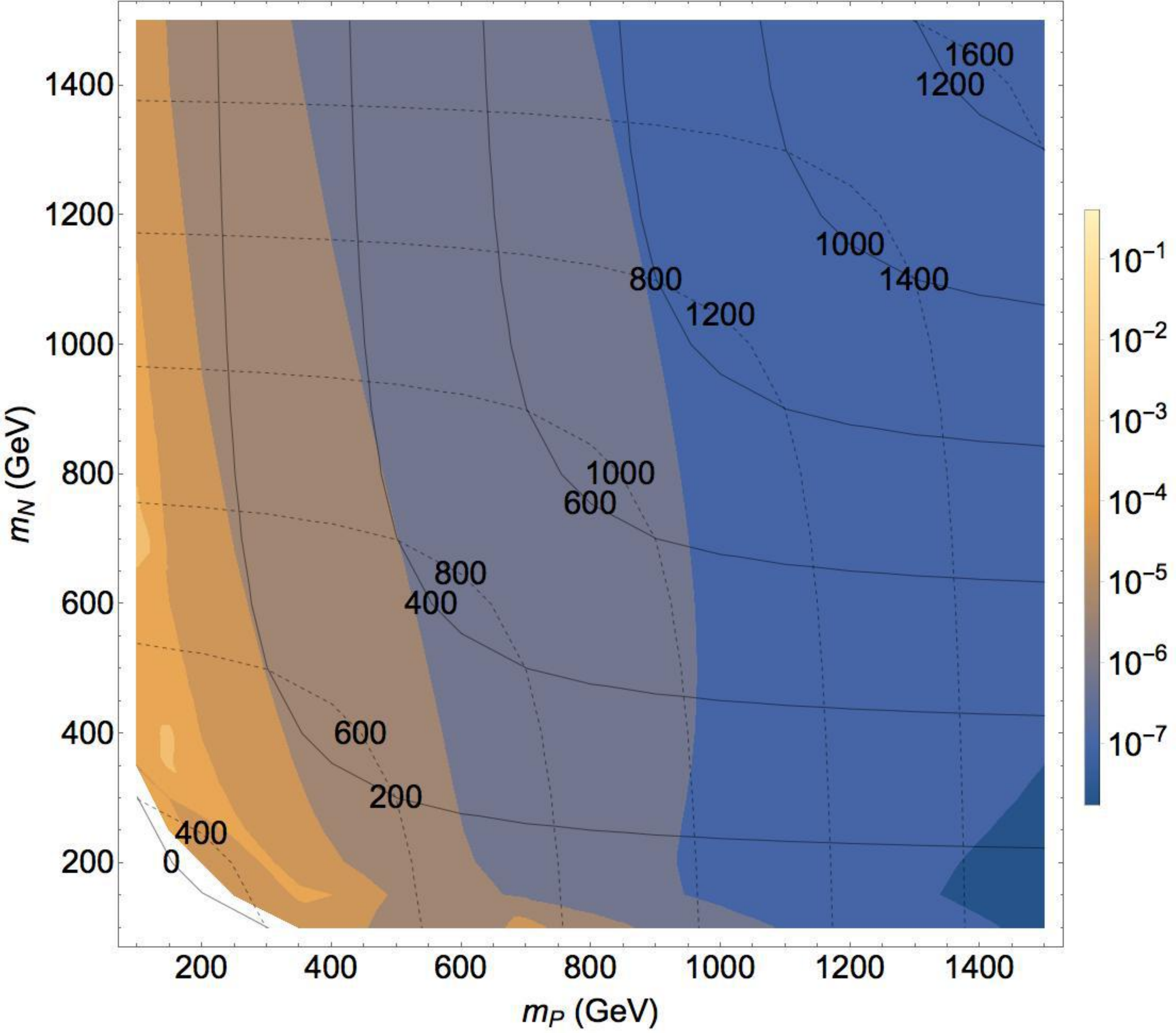}
 \end{center}
 \caption{Branching ratios for $h\to XX$~(left) and $h\to XZ$~(right) decays 
due to mediator fermion loops in the $m_P$-$m_N$ plane for 
$\lambda=1$, $\alpha_x=10\,\alpha$, and $m_x=15\,\gev$. 
The solid~(dashed) black contours indicate $m_1$~($m_2$) masses,
while the solid red line shows the sensitivity of the most sensitive
current LHC searches~\cite{Aad:2015sva}.}
 \label{fig:hbr}
 \end{figure}

  In Fig.~\ref{fig:hbr} we show the range of Higgs branching fractions
for $h\to XX$~(left) and $h\to XZ$~(right) in the $m_P$-$m_N$ plane 
due to fermion loops for $\lambda = 1$, $\alpha_x = 10\alpha$, and
$m_x=15\,\gev$.  The solid~(dashed) lines in the figure 
indicate contours of constant $\psi_1$~($\psi_2$) masses.  
For the $h\to XX$ channel, the branching fractions are symmetric 
in $m_P$ and $m_N$ since both states couple equally to the dark vector. 
The decay fractions for $h\to XZ$ tend to be somewhat lower than for $h\to XX$,
due to weaker effective couplings in the amplitudes for large $\alpha_x$.
In particular, the coupling of either fermion to the $Z$ is at most 
$\bar{g}/2 < \sqrt{10\alpha_x}$ and is maximal for a doublet-like fermion,
while the Higgs coupling relies on mixing between the $P^0$ and $N$ 
gauge eigenstates.  

  These exotic Higgs decay channels to one or more dark vectors also arise
from the standard vector and Higgs portal 
couplings~\cite{Gopalakrishna:2008dv,Curtin:2013fra}.
With only a vector portal, the main new decay is $h\to XZ$ through 
the SM $h\to ZZ$ vertex with one of the $Z$ legs mixing into the dark 
vector $X$~\cite{Falkowski:2014ffa,Curtin:2014cca}.  
The corresponding decay width is suppressed by both $\epsilon^2$
and $(m_x/m_Z)^2$, and tends to have a very small branching fraction 
once other direct constraints on $\epsilon$ are taken 
into account~\cite{Curtin:2014cca}.  The Higgs portal interaction
can lead to $h\to XX$ decays with a significant 
rate~\cite{Gopalakrishna:2008dv,Curtin:2013fra,Curtin:2014cca}.
We have not included the effects of these couplings in the results above.  
For minimal natural values of these parameters in our theory,
we find that their contributions to the Higgs decay amplitudes are much smaller
than those from direct fermion loops over the range of
masses shown in the figure.

\subsection{Experimental Signals}

  Prospects for observing Higgs boson decays to one or more light dark vector
bosons were studied in 
Refs.~\cite{Davoudiasl:2013aya,Gopalakrishna:2008dv,Curtin:2013fra,Falkowski:2014ffa,Curtin:2014cca,Gabrielli:2014oya,Biswas:2015sha,Biswas:2016jsh,Biswas:2017lyg,Campos:2017dgc}.  
If the $X$ vector boson is the lightest state in the hidden sector, 
it decays exclusively to SM final states through its vector portal mixing
with hypercharge.  These decays can have a significant branching fraction 
to charged leptons~\cite{Curtin:2014cca}, typically larger than that of
the $Z$ boson, and are prompt for natural values of
the kinetic mixing $\epsilon$.

  The most recent experimental analysis of rare Higgs decays to dark vectors
is the ATLAS study of Ref.~\cite{Aad:2015sva}, 
based on about $20.5\,\text{fb}^{-1}$ of data at $\sqrt{s} = 8\,\tev$.  
This search uses four-lepton final states with two opposite-sign 
same-flavor~(OSSF) pairs, 
and covers the dark vector mass range $15\,\gev \leq m_x \leq m_h/2$.  
For the $h\to XZ^{(*)}$ channel, the combined invariant mass is required
to reconstruct the Higgs mass to within about $10\,\gev$, and a bump
search is performed on the OSSF lepton pair with the lowest invariant
mass.  Their result can be translated into a limit on the branching
ratio
$\BR(h\to XZ) \lesssim 0.5\!-5\times 10^{-3}$ over the dark vector mass
range covered by the search.  In the $h\to XX$ channel, events with
two OSSF pairs are also selected and grouped such that the resulting
pair of two-body invariant masses are as close as possible.
The exclusion derived corresponds to $\BR(h\to XX) \lesssim 3\times 10^{-4}$
over the vector mass range studied.

Comparing these LHC exclusions to the branching fractions found above due
to the mediator fermions of the vectorized lepton portal, Fig.~\ref{fig:hbr}, 
we find that current data puts a significant limit on the new fermion masses
for $\lambda = 1$ and $\alpha_x=10\alpha$.  Dedicated analyses with the 
full current and expected LHC data sets will have sensitivity to even
larger fermion masses in both the $h\to XX$ and $h\to XZ$ channels.  
Let us also point out that the search of Ref.~\cite{Aad:2015sva}
concentrated on the dark vector mass range of $15\,\gev \leq m_x \leq m_h/2$.
This range is only weakly constrained by direct searches for
dark vectors, with the strongest current bounds coming from precision electroweak
tests that limit $\epsilon \lesssim 0.02$~\cite{Hook:2010tw}.  
The collider sensitivity to smaller dark vector masses is limited by
backgrounds from heavy flavor resonances appearing at masses below
about $11\,\gev$, and from the tendency of the leptons from a lighter
vector boson to be collimated~\cite{Falkowski:2010cm,Falkowski:2010gv}. 
Note, however, that that existing direct limits on light dark vectors are much 
stronger for $m_x \lesssim 11\,\gev$ and constrain 
$\epsilon \lesssim 10^{-3}$~\cite{Alexander:2016aln},
of the same size as the natural range for this coupling in our minimal theory.

  Our analysis shows that exotic Higgs decays to dark vectors from loops
of heavy mediator fermions are potentially observable in future Higgs
searches at the LHC and beyond.  In the sections to follow, we investigate
other constraints on the theory from precision electroweak tests,
direct collider searches, Higgs stability, and dark matter considerations.
In doing so, we set $\lambda =1$, $\alpha_x=10\alpha$, and $m_x=15\,\gev$
as fiducial parameters against which to compare.  For these parameters,
we find that searches for exotic Higgs decay can provide comparable or greater 
sensitivity to the theory than other experimental probes.

\section{Precision Electroweak and Collider Constraints\label{sec:bounds}}  

The vectorized lepton portal can induce significant decay fractions for 
$h\to XX$ and $h\to XZ$ provided $\lambda$ and $\alpha_x$
are relatively large and the vector-like fermions $\psi_1$ and $\psi_2$ 
are not too heavy. In this section we investigate the bounds imposed
on the theory from precision electroweak measurements, direct
collider searches, and Higgs stability.

\subsection{Electroweak Constraints\label{sec:pew}}

  The new $\psi_1$, $\psi_2$, and $P^-$ fermions couple directly to the
electroweak vector bosons, and therefore induce oblique corrections
to precision electroweak observables~\cite{Peskin:1990zt,Peskin:1991sw}.  
In addition, the gauge kinetic mixing of $U(1)_x$ 
with hypercharge leads to mixing between the physical $X$, $Z$, and $\gamma$ 
vector bosons, further modifying these 
observables~\cite{Curtin:2014cca,Hook:2010tw,Babu:1997st,Kumar:2006gm,Chang:2006fp}.  However, for natural ranges of the kinetic mixing parameter 
$\epsilon \lesssim 10^{-2}$ with $m_x \lesssim 30\,\gev$,  
the effects of vector boson mixing are much smaller than current 
limits~\cite{Curtin:2014cca,Hook:2010tw},
and thus we focus exclusively on the effects of the heavy fermions.

  Oblique corrections due to the new fermions are captured effectively
by the Peskin-Takeuchi $S$, $T$, and $U$ parameters.  These have been
computed for vector-like fermions with the same SM quantum
numbers as those considered here in Refs.~\cite{Ellis:2014dza,Chen:2017hak}.  
Full expressions for the corrections to $S$, $T$,
and $U$ are collected in Appendix~\ref{app:pew}.

\begin{figure}[ttt]
 \begin{center}
         \includegraphics[width = 0.45\textwidth]{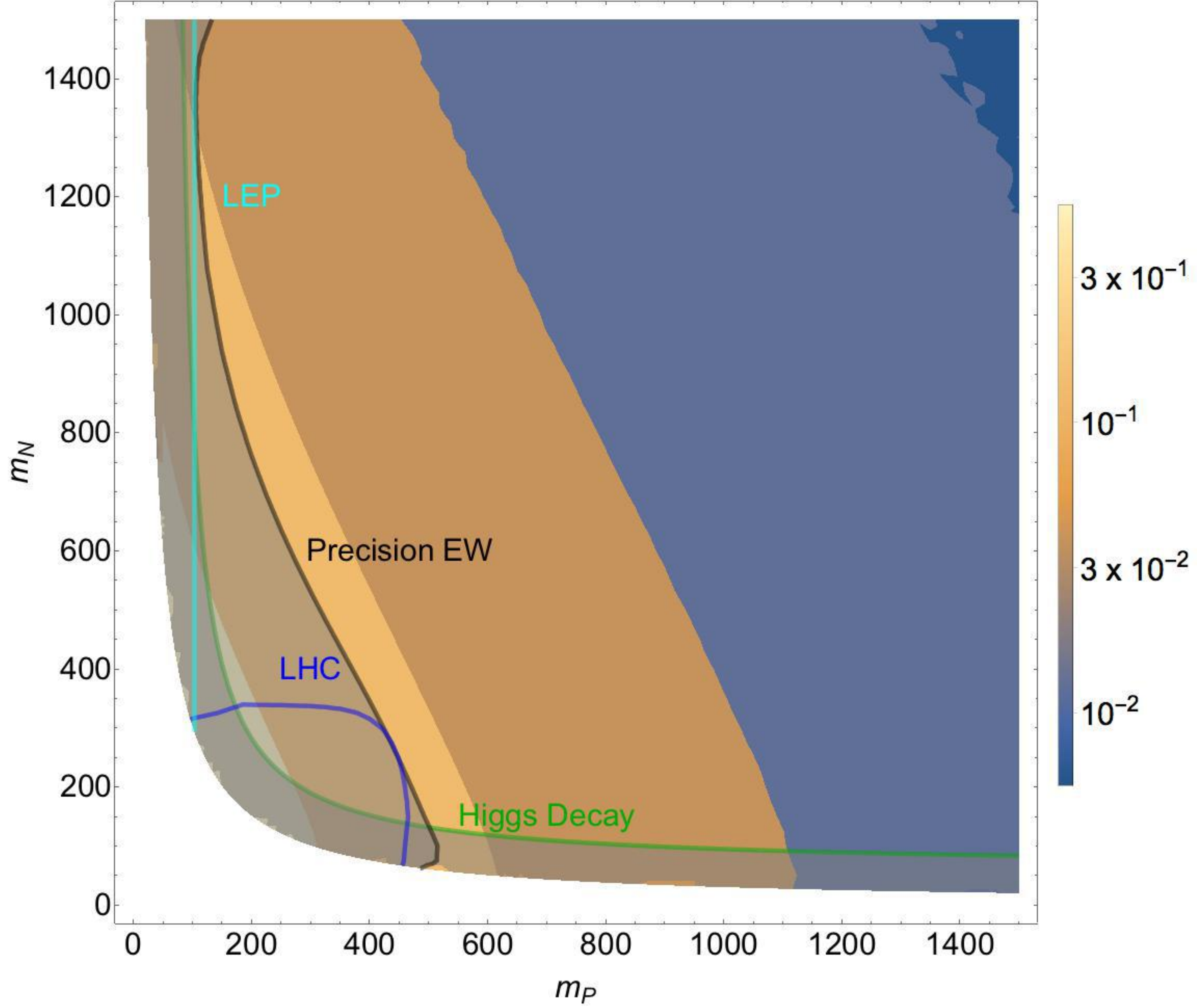}
 \end{center}
 \caption{Precision electroweak and collider constraints on the minimal
vectorized lepton portal for $\lambda = 1$.  
The dark grey line shows the combined exclusion from precision 
electroweak tests, the cyan~(LEP) and blue~(LHC) indicate
bounds from direct collider searches, and the green line shows the limit
from the non-observation of invisible Higgs decays.  The coloured contours
indicate the shifts in the oblique $\Delta T$ parameter due to 
the heavy fermions.}
 \label{fig:pew}
 \end{figure}

To derive an exclusion on the theory from current electroweak data,
we use the central values, uncertainties, and correlations among the
$S$, $T$, and $U$ parameters obtained in the fit of Ref.~\cite{Baak:2014ora}
with $m_t = 173\,\gev$ and $m_h=125\,\gev$.  
The corresponding $95\%~\text{c.l.}$ excluded region in the $m_P$-$m_N$
plane for $\lambda=1$ lies to the left of the solid black line 
in Fig.~\ref{fig:pew}.  We find that the corrections to $S$ and $U$ from 
the new fermions are typically very small, and the primary effect of
the fermions is to shift the $T$ parameter, related to the mass splitting 
of the components of the electroweak doublet $P$ from mixing with $N$.  
Contours of $\Delta T$ are also shown in Fig.~\ref{fig:pew}, 
and the excluded region is approximated well by the condition 
$\Delta T \lesssim 0.14$.

\subsection{Collider Bounds}

  Collider searches for the charginos and neutralinos of supersymmetry 
can be applied to the vector-like fermions we 
are considering.  In particular, our system consists of an electroweak 
doublet and singlet, and is similar in its collider phenomenology to 
a Higgsino-Bino system~\cite{Cohen:2011ec}.
The lightest new fermion in the theory is $\psi_1$, which is stable
and contributes to missing energy in analogy to the lightest $\chi_1^0$ 
neutralino.  We estimate here the limits on the $\psi_1$, $\psi_2$, 
and $P^-$ massive fermions by reinterpreting searches 
for electroweak superpartners at LEP~II and the LHC.

    For $m_P \ll m_N$, the lighter $\psi_1$ and $P^-$ states both come 
mainly from the electroweak doublet and tend to be fairly close in mass, 
similar to a set of light Higgsinos ($\mu \ll M_1,\,M_2$).
In contrast to light Higgsinos, however, the neutral state is a single
Dirac fermion $\psi_1$ instead of a pair of Marjorana $\chi_1$ and $\chi_2$
modes.  The charged $P^-$ state is similar to the lightest chargino $\chi_1^+$
in its production, with decays through $P^-\to W^{-(*)}\psi_1$.  
The heaviest state $\psi_2$ is mostly singlet, and will therefore have
suppressed production through electroweak vector bosons.
With $m_P \ll m_N$, it decays in a roughly $2\!:\!1\!:\!1$ proportion via
$\psi_2\to W^+P^-$, $\psi_2\to Z\psi_1$, 
and $\psi_2\to h\psi_1$~\cite{Kumar:2015tna}.

  In the opposite limit, $m_P \gg m_N$, the lightest state $\psi_1$ 
is mostly singlet while the heavier $P^-$ and $\psi_2$ particles 
are Higgsino-like.  They decay via $P^-\to W^-\psi_1$, along with
$\psi_2\to Z\psi_1$ and $\psi_2\to h\psi_1$ in a roughly $1\!:\!1$ 
ratio~\cite{Kumar:2015tna}.
This system is similar to the electroweakino sector of a supersymmetric
theory with $M_1 \ll \mu \ll M_2,\,m_{sfermion}$.  

  Searches for superpartners at LEP~II are summarized in Ref.~\cite{lepsusy}.  
The most relevant channels for our scenario are the chargino modes
$e+e^-\to \chi_1^+\chi_1^-$ with $\chi_1^+ \to \chi_1^0\,W^{+(*)}$~\cite{lepchg1}.
These can be applied directly to $P^+P^-$ production. 
For $\lambda = 1$ and $m_N \lesssim 2\,\tev$, the chargino limits translate into
\beq
m_P > 103\,\gev \ .  
\eeq
This value of $\lambda$ (and the condition $\sqrt{m_Nm_P} > \lambda v$
mentioned in Sec.~\ref{sec:setup}) 
also implies that $(m_1+m_2)$ is always larger than the maximal LEP~II 
center-of-mass energy, so no bounds are obtained from searches for 
$\chi_1^0\chi_2^0$ production in this case.  Searches for neutralino LSPs with
initial-state photon radiation can also be applied to 
$e^+e^-\to \psi_1\overline{\psi}_1\gamma$~\cite{Fox:2011fx}, 
but we find production 
cross sections well below the limit from Ref.~\cite{Abdallah:2003np}.

  More recently, the LHC collaborations have extended the constraints on
electroweak superpartners to masses beyond the reach of LEP~II.  
The new fermions in our theory can be produced by electroweak Drell-Yan channels,
through an off-shell Higgs 
boson~\cite{Buckley:2014fba,Harris:2014hga,Craig:2014lda}, and via an
$s$-channel dark vector.  Whenever the production cross section is large
enough to be potentially observable, we find that it is dominated
by standard Drell-Yan (for fiducial values of $\lambda=1$ and $\epsilon=10^{-3}$).
Decays of the heavier fermions to the stable $\psi_1$ produce signals with
jets, leptons, and missing energy in direct analogy to supersymmetric cascades.
Radiation of dark vectors by these fermions can produce
additional visible objects in the 
events~\cite{Strassler:2006im,Strassler:2006qa,Han:2007ae}.   
We do not expect such radiation to have a significant qualitative effect
on the searches considered here that rely mainly on leptons that 
reconstruct a $Z$ boson, but they could open new search channels at the LHC.

  The most constraining LHC search for our theory appears to be
the CMS opposite-sign same-flavor~(OSSF) dilepton analysis 
of Ref.~\cite{CMS:2017vwf}.
This search was based on $35.9~\text{fb}^{-1}$ of data at a center-of-mass
energy of $\sqrt{s} = 13\,\tev$.  The channels in the analysis relevant 
for our theory were those designed for $\chi_2^0\,\chi_1^{\pm}$
production followed by $\chi_2^0\to \chi_1^0\,Z$ 
and $\chi_1^{\pm}\to \chi_1^0\,W^{\pm}$.  These channels required exactly
two isolated light-flavor OSSF leptons with 
$86\,\gev < m_{\ell\ell} <96\,\gev$, 
at least two jets with $p_T> 35\,\gev$ and $m_{jj} < 110\,\gev$, 
and missing energy $\met > 100\,\gev$.  Vetoes on additional leptons
and $b$-tagged jets were also applied.  These channels will receive
contributions from $\overline{\psi}_2\,P^{-}$ and $\psi_2\,\overline{\psi}_2$ 
production with one on-shell $\psi_2\to Z\psi_1$ decay.  

  To estimate exclusion limits from this search, we use
\texttt{MadGraph5\_aMC@NLO}~\cite{Alwall:2014hca} with a model implemented in \texttt{FeynRules}~\cite{Alloul:2013bka}
to compute the relevant LHC production cross sections at $\sqrt{s}=13\,\tev$.  
We then compare to the cross section limits obtained in 
Ref.~\cite{CMS:2017vwf} for a Wino-like $\chi_2^0\chi_1^{\pm}$ simplified 
model in which the two electroweakino states are degenerate and assumed 
to decay exclusively to a stable $\chi_1^0$ state through the weak vector bosons.
In making the comparison, we include $P^{-}\overline{\psi_2}$,
$P^+\psi_2$, and $\psi_2\overline{\psi}_2$ production, and we rescale 
their cross sections by the branching fraction for $\psi_2\to Z\psi_1$.
The main simplification we make in deriving our exclusions is 
the assumption that the detection efficiencies are approximately 
the same in our theory as for the simplified electroweakino model.
We also take the exclusion cross section to be 
$\sigma_{tot} < 0.01\,\text{pb}$.  Both assumptions are somewhat agressive,
and thus we expect our result to represent an upper limit on the exclusion
derived from a full recasting of the CMS search.  Our result is
shown in Fig.~\ref{fig:pew}.

  Other potentially relevant LHC searches are the trilepton analysis
of Ref.~\cite{CMS:2017fdz} and the mass-degenerate dilepton analysis
of Ref.~\cite{CMS:2017fij}.  Comparing their excluded cross sections
to those of our theory, we do not find any limits beyond the dilepton
analysis described above.

\subsection{Higgs Stability\label{sec:stability}}

  New fermions with large Yukawa couplings to the Higgs field
can destabilize the Higgs potential.  They do so by modifying the
renormalization group~(RG) evolution of the Higgs self coupling
$\lambda_H$, and tend to drive it negative at a lower
scale than in the SM~\cite{Kribs:2007nz,Hashimoto:2010at,Ishiwata:2011hr,ArkaniHamed:2012kq}.  
Without additional new physics near the scale at which this occurs,
the tunnelling rate from the standard electroweak vacuum to the unstable region
at large Higgs field values tends to be shorter than the age of 
the universe~\cite{Isidori:2001bm,Buttazzo:2013uya}.  At best, 
this instability can be taken to be an upper cutoff for the consistency 
of the theory.

  To investigate these effects, we evolve the couplings of the theory
to higher scales using the one-loop RG equations for the system.
These are listed in Appendix~\ref{app:rg}, and generalize the results 
of Refs.~\cite{Machacek:1983tz,Machacek:1983fi,Machacek:1984zw}.
As inputs, we use the $\overline{MS}$ values for the relevant SM parameters
derived in Ref.~\cite{Martin:2016xsp} defined at scale $\mu_t = 173.34\,\gev$:
\beq
\begin{array}{rclrclrcl}
g_1 &=& \sqrt{5/3}\,(0.3585) \ ,~~~~~&g_2&=& 0.6476 \ ,~~~~~&g_3 &=& 1.1667 \ ,\\
y_t &=& 0.9369 \ ,~~~~~&\lambda_H &=& 0.12597 \ .
\end{array}
\label{eq:rginput}
\eeq
These inputs are evolved up to the fiducial massive fermion scale 
$\mu_F = 500\,\gev$ as in the SM, and then from $\mu_F$ to higher scales
in the full theory with heavy fermions and the dark vector boson.

  As expected, we find that the new Yukawa coupling $\lambda$ drives
the Higgs quartic coupling $\lambda_H$ negative more quickly than in the SM.
The condition we apply for the metastability of the standard electroweak
vacuum follows Ref.~\cite{ArkaniHamed:2012kq}, which is based
on Ref.~\cite{Isidori:2001bm},
\beq
\lambda_H(\Lambda_H) = -0.065\left[1- 0.02\ln(\lambda_H/\mu_t)\right] \ .
\label{eq:meta}
\eeq
This relation defines $\Lambda_H$, the maximum scale at which new physics
that stabilizes the Higgs potential must emerge.  Numerically, $\Lambda_H$
tends to be one or two orders of magnitude larger than the scale at which
the Higgs quartic coupling $\lambda_H$ runs negative~\cite{Buttazzo:2013uya}.

  For the inputs listed in Eq.~\eqref{eq:rginput} together with 
$\lambda = 1$ and $\alpha_x=10\alpha$ at $\mu_t$, we find a Higgs instability
cutoff scale of $\Lambda_H \simeq 4.6\times 10^4\,\gev$, with the quartic coupling
running negative at $\mu \simeq 1.0\times 10^4\,\gev$.
There is also a Landau pole in the new gauge coupling at scale 
$\mu \simeq 10^{11}\,\gev$ 
for $\alpha_x(\mu_t) = 10\alpha$.  Reducing $\alpha_x(\mu_t)$
quickly pushes up the scale at which the Landau pole occurs, but has only
a mild (lowering) effect on $\Lambda_H$.  The Higgs instability scale
for $\lambda=1$ is relatively low, but is still high enough to justify 
our treatment of the heavy fermions provided we interpret the theory 
as an effective one with a cutoff near $5\,\tev$.  
Even so, we note that $\lambda=1$ is close 
to the upper limit of what is possible for the consistency of 
our previous analyses.\footnote{There is a significant sensitivity 
of these results to the SM input parameter values for small $\lambda \ll 1$
reflecting a theoretical uncertainty on our one-loop treatment.  
However, for $\lambda \sim 1$, the new Yukawa coupling dominates
and the dependence on the SM inputs becomes modest.}

\section{Connections to Dark Matter\label{sec:dm}}

  In the minimal realization of the vectorized lepton portal, the lightest
exotic fermion $\psi_1$ is stable and contributes to the density 
of dark matter~(DM).  This state is also a Dirac fermion with direct couplings
to the $Z^0$ and $X$ vector bosons, implying that it can have a 
large spin-independent scattering cross section with nuclei.  
We investigate these features in this section and show that they 
impose strong constraints on the model assuming standard thermal 
production of $\psi_1$ in the early universe.  These constraints 
can be evaded in scenarios with low effective reheating temperatures 
or by going beyond the minimal realization of the theory.

\subsection{Relic Densities}

  The relic density of $\psi_1$ particles from thermal freeze-out
is determined by its dominant annihilation cross sections
to dark vectors, electroweak vectors, and Higgs final states.
Annihilation to pairs of dark vector bosons in our scenario is identical
to minimal models of secluded dark matter~\cite{Pospelov:2007mp}, 
with leading cross section
\beq
\langle\sigma v\rangle_{XX} ~\simeq~ 
\frac{\pi\alpha_x^2}{m_1^2}\sqrt{1-\lrf{m_x}{m_1}^2}
\ .
\eeq
The complete expression can be found in Ref.~\cite{Cline:2014dwa}.
Since both the $N$ and $P$ states couple in the same way to the dark vector,
this cross section is independent of their mixing, and depends
only on the gauge coupling $\alpha_x$ and the mass $m_1$.\footnote{
A light dark vector coupled to heavier dark matter can also enhance 
the annihilation cross section (in all channels) by the Sommerfeld 
effect~\cite{Baer:1998pg,Lattanzi:2008qa,Falkowski:2009yz}.  
We find that this enhancement is very mild for $\alpha_x \leq 10\alpha$
and $m_x=15\,\gev$.}

  For direct annihilation to SM final states, the most important modes are
typically $\psi_1\overline{\psi_1}\to ZZ,\,WW,\,hh$.  
These cross sections depend sensitively on the mixing between the 
$P^0$ and $N$ gauge eigenstates that combine to make up $\psi_1$ and $\psi_2$.  
The $s$-wave amplitude for the $WW$ channel is facilitated by a $t$-channel 
$P^{-}$ exchange and scales proportionally to $s_{\alpha}^{2}$, while
the analogous $ZZ$ process involves $t$-channel $\psi_{1}$ or $\psi_{2}$ 
exchange and is proportional to $s_{\alpha}^{4}$ or $s_{\alpha}^{2}c_{\alpha}^{2}$ 
respectively. In the event of small mixing angles, $p$-wave processes 
involving an $s$-channel Higgs can be significant.  These amplitudes
scale as $\lambda s_{\alpha}c_{\alpha}$, and include $WW$, $ZZ$,
or $hh$ final states.

\begin{figure}[ttt]
 \begin{center}
         \includegraphics[width = 0.42\textwidth]{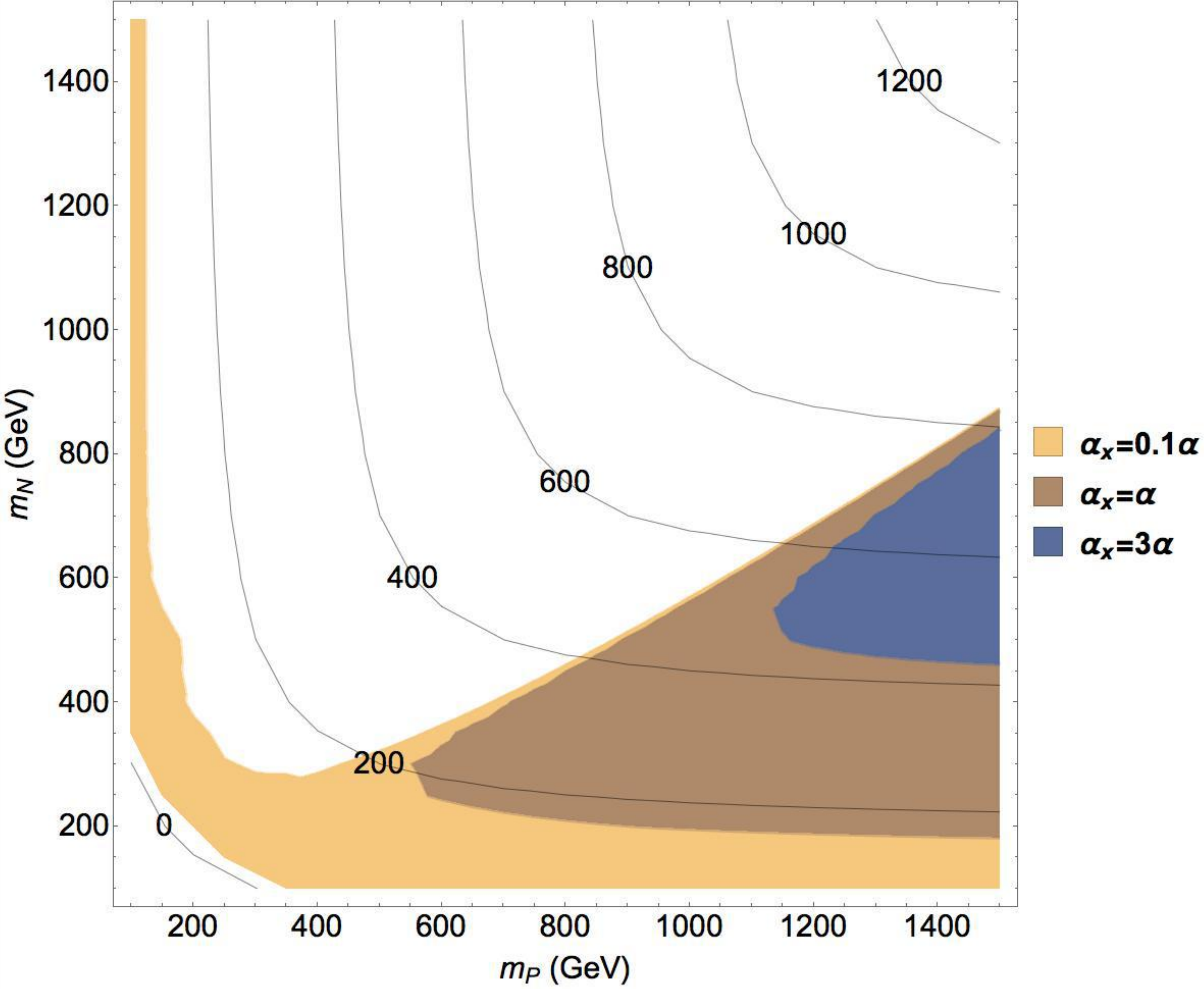}~~~~~~~
				 \includegraphics[width = 0.42\textwidth]{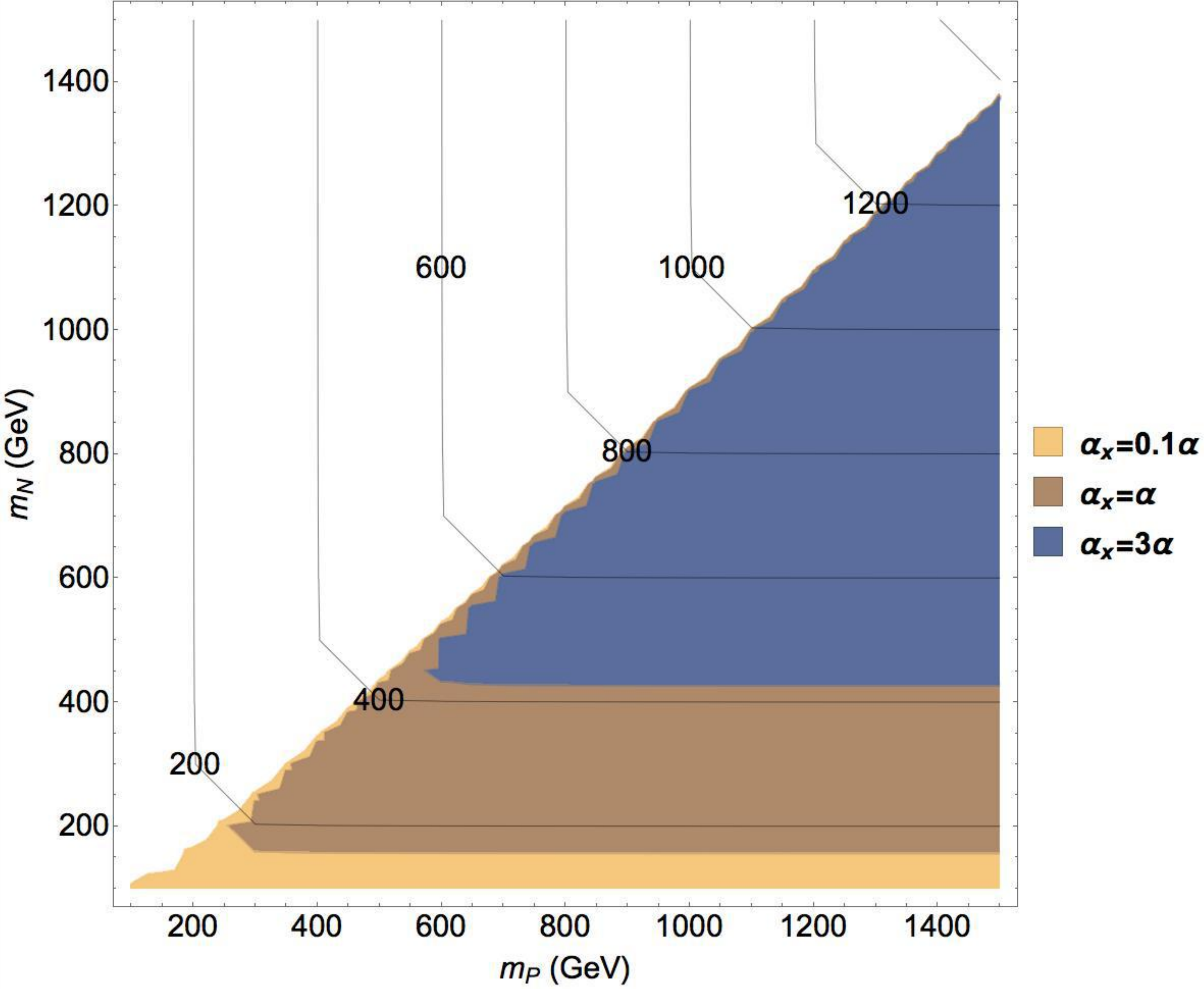}
 \end{center}
 \caption{Regions in the $m_P$-$m_N$ where the thermal $\psi_1$ relic density
exceeds the observed value. The plot on the left has $\lambda = 1$, 
while the plot on the right corresponds to $\lambda = 0.1$. 
The different shadings show the exclusions for $\alpha_x = 0.1\alpha,\,\alpha,\,3\alpha$.  Also shown are contours of the $\psi_1$ mass $m_1$.}
 \label{fig:reld}
 \end{figure}

  In Fig.~\ref{fig:reld} we show regions in the $m_{P}\!-\!m_{N}$ plane
where the $\psi_1$ relic density exceeds the observed value.
In this figure, we set $\lambda = 1$~(left) and $\lambda=0.1$~(right), 
with $m_x= 11\,\gev$ and several values of 
$\alpha_x = 0.1\alpha,\,\alpha,\,3\alpha$.  Setting $\alpha_x=10\alpha$,
the annihilation to dark vectors becomes very efficient and the entire
parameter region shown yields an acceptable relic density.
Also shown are contours of the $\psi_1$ mass $m_1$.


\subsection{Direct Detection}

 Direct searches for DM scattering put strong bounds
on spin-independent DM-nucleon effective cross sections, 
on the order of $\sigma_{SI} \lesssim 10^{-46}\,\text{cm}^2$
for $m_{DM} \sim 100\,\gev$.  This is orders or magnitude
larger than the effective per-nucleon cross section of a
stable Dirac fermion with electroweak charge,
$\sigma_{SI} \simeq 10^{-39}\,\text{cm}^{-2}$~\cite{Jungman:1995df}.  
As a result, current direct detection bounds are sensitive to Dirac fermion 
relics that make up only a tiny fraction of the full dark matter 
density~\cite{Halverson:2014nwa}.

  The spin-independent nucleon cross section of the $\psi_1$ state
receives contributions from $Z$, $X$, and Higgs exchange.
The corresponding effective operators have the same non-relativistic
limit and interfere with each other.  Together, they imply an effective 
per-nucleon cross section of~\cite{Jungman:1995df}
\beq
\sigma_{SI} = \frac{\mu_{n}^{2}}{\pi}\left[\frac{f_pZ+f_n(A-Z)}{A}\right]^2 \ ,
\label{eq:sigsi}
\eeq
where $\mu_n$ is the DM-nucleon reduced mass, $A$ and $Z$ describe the 
target nucleus, and 
\beq
f_p &=& \frac{G_F}{\sqrt{2}}s_{\alpha}^2(1-4s_W^2)
~-~ \frac{4\pi}{m_x^2}\,\epsilon\,q_x\,\sqrt{\alpha\,\alpha_x}
~+~ {\tilde{d}_p}\left[\frac{2}{9}+\sum_{q}f_{T,q}^{p}\right] \ ,
\label{eq:fp}\\
f_n &=& -\frac{G_F}{\sqrt{2}}s_{\alpha}^2
~+~ 0
~+~ {\tilde{d}_n}\left[\frac{2}{9}+\sum_{q}f_{T,q}^{n}\right] \ .
\label{eq:fn}
\eeq
In both expressions above, the first term is due to $Z$ exchange,
the second to $X$ exchange, and the third to Higgs exchange.
The $X$ exchange terms depend on the sign of the dark charge
of $\psi_1$ and assume $m_x \gtrsim 100\,\mev$.  For the Higgs exchange
terms, the quantity $\tilde{d}_{p,n}$ is given by
\beq
\tilde{d}_{p,n} = \frac{m_{p,n}}{v}\,\frac{\lambda\,s_{\alpha}c_{\alpha}}{m_h^2} \ ,
\eeq
where the sums run over $q=u,d,s$, and the coefficients
$f^{N}_{T,q}$ can be found 
in Refs.~\cite{Hill:2014yxa,Bishara:2016hek,Junnarkar:2013ac}.

Combining these expressions with the relic densities calculated previously,
we find density-weighted per-nucleon cross sections,
$(\Omega_1/\Omega_{DM})\sigma_{SI}$, that are typically much larger
than the current constraints from PandaX~\cite{Tan:2016zwf} 
and LUX~\cite{Akerib:2016vxi}.  This applies even for $\alpha_x=10\alpha$
when the $\psi_1$ relic density is significantly smaller than the total dark 
matter density.  Dark photon exchange dominates for smaller 
$m_x$ and natural one-loop values of the kinetic mixing $\epsilon$.
Even with this contribution suppressed by $\epsilon\to 0$,
the scattering due to $Z$ exchange still tends to be too large.

\begin{figure}[ttt]
 \begin{center}
         \includegraphics[width = 0.44\textwidth]{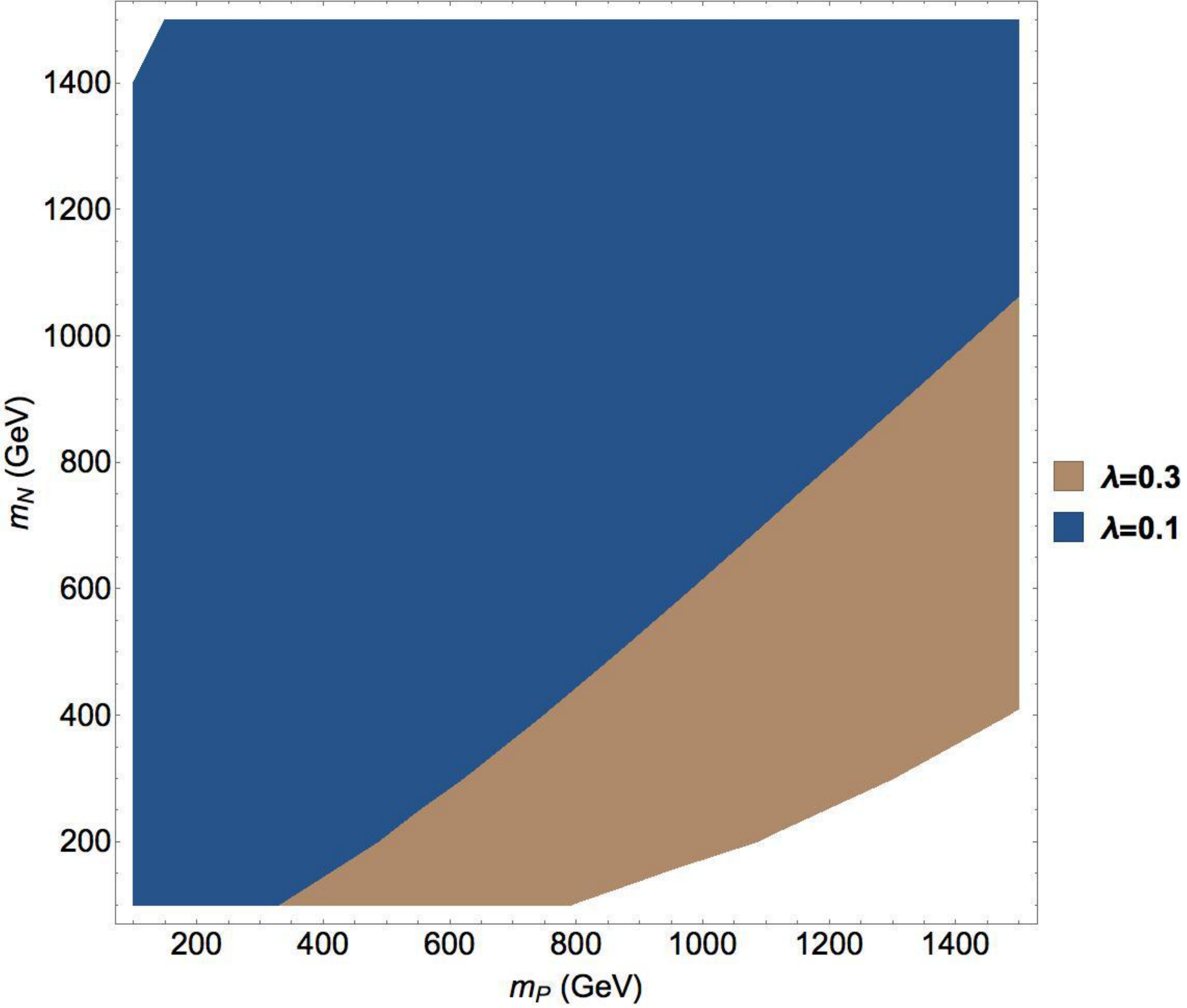}~~~~~
         \includegraphics[width = 0.423\textwidth]{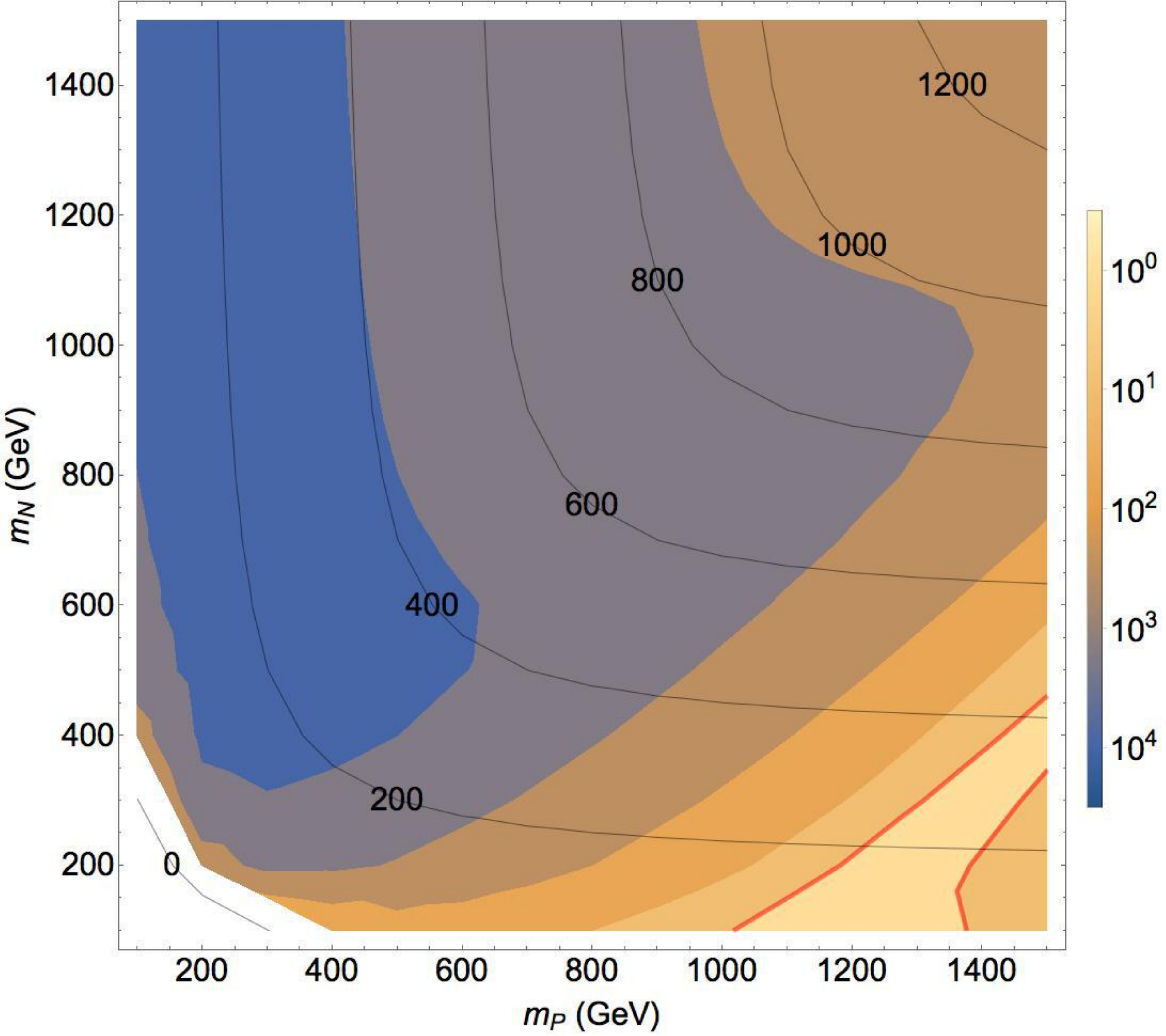}			
 \end{center}
 \caption{Two examples of how a thermal $\psi_1$
relic density with $\alpha_x=10\alpha$ can be consistent with the current 
LUX bounds on spin-independent dark matter scattering~\cite{Akerib:2016vxi}.
The plot on the left illustrates the small $\lambda$ scenario, in which
$\epsilon\to 0$, the lighter $\psi_1$ state is very singlet-like with
suppressed mixing with the doublet.  The unshaded regions at the lower
right for $\lambda=0.1,\,0.3$ where the $\psi_1$ state is mostly singlet
are allowed by current bounds.  The right panel illustrates the scenario
where the contributions to nucleon scattering from the
dark photon and the $Z$ boson cancel against each other.  This plot shows
contours of $(\Omega_1/\Omega_{DM})\sigma_{SI}/\sigma_{LUX}$ for $\lambda = 1$, 
$\epsilon = 10^{-4}$, and $m_x=10\,\gev$.  The allowed region between
the solid red lines.  
%
}
 \label{fig:dd}
 \end{figure}

  Two potential loopholes to these bounds exist.  The first requires
a very small $\epsilon \to 0$ to suppress dark photon exchange
together with a lighter singlet-like $\psi_1$ state to reduce the $Z$
and Higgs contributions to nucleon scattering.  A large value of $\alpha_x$ 
is also needed to yield a small $\psi_1$ relic density.  While these
parameter values can give acceptably small density-weighted cross sections,
they correspond to $s_{\alpha} \sim \lambda v/m_P \ll 1$ and imply
a strong suppression of Higgs decays to dark photons.
This is illustrated in the left panel of Fig.~\ref{fig:dd}, where we show
the parameter regions excluded by LUX~\cite{Akerib:2016vxi} for 
$\lambda=0.1,\,0.3$, $\epsilon\to 0$, and $\alpha_x=10\alpha$.
The unshaded regions at the lower right are allowed.

The second loophole arises when there is a strong cancellation
between the dark photon and $Z$ boson contributions to the cross section.
Suppression of the cross section from such a cancellation
is limited by the mixture of isotopes present in natural xenon
to about $2\times 10^{-4}$ relative to $f_p=f_n$.  Moreover, the optimal
suppression for xenon is different from that for other materials
such as the germanium used in CDMS-II~\cite{Ahmed:2009zw}. 
The allowed region of parameter space in this context for $\lambda=1$,
$\epsilon = 10^{-4}$, $m_x=10\,\gev$, and $\alpha_x=10\alpha$ is
illustrated in the right panel of Fig.~\ref{fig:dd}, where we show contours 
of the density weighted spin-independent cross section relative to the bound from
LUX~\cite{Akerib:2016vxi}, $(\Omega_1/\Omega_{DM})\sigma_{SI}/\sigma_{LUX}$.
The region between the solid red lines is consistent with current limits. 

\subsection{Beyond the Minimal Scenario}

  Our analysis shows that the lightest $\psi_1$ fermion is very strongly
constrained by dark matter direct detection, particularly when the 
Higgs branching fraction to dark vectors is significant.  
A similar conclusion was obtained in Ref.~\cite{Davoudiasl:2012ig}.  
These constraints can be avoided if there is non-thermal cosmological evolution
or additional structure in the theory.

  The relic densities used in making the estimates above assumed thermal 
cosmological evolution during and after the freezeout of the stable 
$\psi_1$ state.  Much smaller relic densities can arise from non-thermal 
evolution.  For example, late reheating following a period of inflation 
or matter domination with a reheating temperature below the 
freeze-out temperature of $\psi_1$ can yield a relic density that is orders
of magnitude below the thermal value~\cite{Gelmini:2006pw}.  
Even so, the tiny remaining abundance of $\psi_1$ could still be 
observable in direct detection experiments due to its large 
spin-dependent scattering cross section~\cite{Halverson:2014nwa}.

  Constraints on the $\psi_1$ abundance from direct detection
can also be reduced if it obtains a small Majorana mass or is able to
decay~\cite{Davoudiasl:2012ig}.  A Majorana mass for $\psi_1$ can arise from
the dark Higgs coupling listed in Eq.~\eqref{eq:maj}.  Such a mass
term will split the four-component $\psi_1$ Dirac fermion into a pair 
of Majorana fermions, and thereby remove the dominant 
contribution to spin-independent elastic scattering from vector boson 
exchange~\cite{TuckerSmith:2001hy}.  The residual vector-mediated 
inelastic scattering is highly suppressed for mass splitting above 
about $\Delta m_1 > 200\,\text{keV}$.  We find that the remaining 
spin-independent scattering due to Higgs excange can still be significant
for $\lambda = 1$, but it can lie below current limits for the subleading 
$\psi_1$ relic densities that occur for $\alpha_x=10\alpha$.  
Alternatively, the operator of Eq.~\eqref{eq:lcoup} allows the $\psi_1$
state to decay to SM fermions through an electroweak vector boson,
in which case the limits from dark matter searches are not relevant.
While the coupling of Eq.~\eqref{eq:lcoup} is constrained by searches
for lepton flavor violation, it is not difficult to avoid these limits
while ensuring that $\psi_1$ decays occur before the onset of
primordial nucleosynthesis.

\section{Comments on the Non-Abelian Case\label{sec:nonab}}

  The vectorized lepton portal can also connect the SM to non-Abelian 
dark gauge groups.  This arises in some theories addressing electroweak 
naturalness~\cite{Burdman:2006tz,Cai:2008au}, typically with a dark 
gauge group of $G_x = SU(3)$, 
but the general structure can emerge more 
broadly~\cite{Juknevich:2009gg,Faraggi:2000pv,Juknevich:2009ji,Boddy:2014yra,Boddy:2014qxa}.  These more general groups can produce important changes
in experimental observables compared to $G_x=U(1)_x$.   
While many of these effects have been discussed in other contexts, 
we review them briefly here and point out a few particular
features of our minimal construction.  To be concrete, 
we focus here on $G_x =SU(N_x)$ with $P$ and $N$ transforming 
in a complex representation $r$, as discussed 
in Ref.~\cite{Juknevich:2009gg,Juknevich:2009ji}.

\subsection{Higgs Decays to Dark Glueballs}
 
  Instead of dark photons, the new vector bosons will be analagous to gluons.  
If there is no symmetry breaking in the dark sector and no other matter fields, 
the \emph{dark gluons} will confine to form \emph{dark glueballs} 
at the scale $\Lambda_x$.  Here and for the rest of this section,
we assume $\Lambda_x \ll m_h$ so that the direct effects of the new fermions
can be treated in perturbation theory, and we define $\alpha_x$ to be the running 
dark coupling at scale $\mu = m_h$.  The dark confinement scale is 
approximately
\beq
\Lambda_x \simeq m_h\,
\exp\left(-\frac{6\pi}{11N_x\alpha_x}\right) \ .
\eeq
This falls very quickly with decreasing $\alpha_x$: for $N_x=3$ we find
$(\alpha_x,\Lambda_x) \simeq (15\alpha,\,1\,\gev)$, $(10\alpha,\,75\,\mev)$, 
$(6.2\alpha,\,1\,\mev)$, and $(\alpha,\,10^{-27}\,\mev)$.

The lightest dark glueball has quantum numbers $J^{PC} = 0^{++}$
and mass (for $N_x=3$) 
$m_0 \simeq 6.8\Lambda_x$~\cite{Morningstar:1999rf,Chen:2005mg}, 
but several other metastable glueballs arise as well.
The effective Higgs interaction induced by the $N$ and $P$ fermions
can be obtained by generalizing the calculation of Sec.~\ref{sec:higgs}:
\beq
-\lag_{eff} \supset \frac{\alpha_x\,T_2(r)}{6\pi}\frac{\lambda^2}{m_1m_2}\,
H^{\dagger}H\,X^{a}_{\mu\nu}X^{a\,\mu\nu} \ .
\label{eq:hop}
\eeq
After confinement and electroweak symmetry breaking, this operator
induces a Higgs portal coupling between the $0^{++}$ glueball and the
SM Higgs boson, allowing it to decay with width~\cite{Juknevich:2009gg}
\beq
\Gamma_{0^{++}} ~\simeq~ 
\left[\frac{T_2(r)}{6\pi}\frac{\lambda^2}{m_1m_2}\right]^2
\left|\frac{\sqrt{2}v\,F_S}{m_0^2-m_h^2-i\Gamma_hm_h}\right|^2
\Gamma_h(m_0)
\ ,
\eeq
where $F_S \simeq (N_x/3)2.3\,m_0^3$ is a glueball matrix element
determined on the lattice~\cite{Chen:2005mg,Meyer:2008tr}, and $\Gamma_h(m_0)$
is the decay width the SM Higgs would have if its mass were equal
to $m_0$.  Like the confinement scale, the glueball decay width varies
extremely rapidly with the value of the running dark gauge coupling at $m_h$.
Setting $\lambda = 1$ and $\sqrt{m_1m_2} = 500\,\gev$,
we find a lifetime of $\tau \simeq 1\,\text{s}$ for $\alpha_x(m_h) = 12\,\alpha$,
and a decay length of $c\tau = 1\,\text{mm}$ for $\alpha_x(m_h) = 23\,\alpha$.\footnote{Fermion loops also yield dimension-eight operators connecting the dark gluons to SM vector bosons, but these yield much smaller decay widths for the parameter ranges of interest~\cite{Faraggi:2000pv,Juknevich:2009ji}.}

  Higgs decays to dark glueballs also proceed through the operator 
of Eq.~\eqref{eq:hop}~\cite{Craig:2015pha,Curtin:2015fna}.  
For light glueball masses, $m_0 \ll m_h/2$,
the inclusive glueball branching fraction follows
that for decays to dark photons up to a simple rescaling:
\beq
\text{BR}(h\to \text{glueballs}) 
~\simeq~
\text{BR}(h\to XX)\times\frac{T_2^2(r)(N_x^2-1)}{q_x^4} \ .
\label{eq:hbrglu}
\eeq
When $m_0$ approaches $m_h/2$, resonances in the final state can modify
the branching fraction in important ways~\cite{Craig:2015pha,Curtin:2015fna}.
Note as well that there is no $h\to (Z+\text{glueballs})$ decay channel 
in the absence of gauge symmetry breaking in the dark sector.
The glueball final states from Higgs decays tend to be long-lived
and appear as simple missing energy unless $\alpha_x(m_h)$ is much larger 
than $\alpha$.  For moderate $\alpha_x(m_h)$, dedicated far detectors 
at the LHC could be sensitive to very late glueball decays~\cite{Chou:2016lxi}.
Very large values of $\alpha_x(m_h)$ can give rise to displaced decays 
within ATLAS or CMS~\cite{Craig:2015pha,Curtin:2015fna}, or produce
emerging or semivisible jets~\cite{Schwaller:2015gea,Cohen:2015toa}.

  Other dark vector decay modes can arise when there is symmetry breaking
in the dark sector above the confinement scale.  For example, 
an adjoint dark Higgs field with a Yukawa coupling $\xi$ to 
the $N$ or $P$ fermions gives rise to the operator
\beq
\mathcal{O} ~\sim~ \frac{\sqrt{\alpha_x\alpha}\,\xi\,T_2(r)}{4\pi}\frac{1}{m}
\Phi^aX_{\mu\nu}^aB^{\mu\nu} \ ,
\eeq
where $m_{\psi} \sim m_1,\,m_2$.  This produces a kinetic
mixing interaction for $\Phi^a\to \langle\Phi^a\rangle$,
and could allow more rapid decays of (some of) the dark 
vector bosons~\cite{ArkaniHamed:2008qn,Baumgart:2009tn,Choquette:2015mca,Barello:2015bhq}.

\subsection{Constraints}

  Bounds from precision electroweak tests and
Higgs stability are mostly independent of the low-energy dynamics
of the dark sector.  The shifts in the oblique parameters discussed
in Sec.~\ref{sec:pew} are enhanced by a factor of $d(r)$,
the dimension of the $G_x$ representation of $N$ and $P$.
For $G_x = SU(3)$ with $r = \mathbf{3}$ and $\lambda=1$, 
this leads to an exclusion of $m_P \gtrsim 1000\!\,-\,\!400\,\gev$
for $m_N = 0\!\,-\,\!1500\,\gev$. 

  The renormalization group equations relevant for a Higgs stability
analysis with a general non-Abelian group $G_x$ and fermion representation $r$
are collected in Appendix~\ref{app:rg}.  For a given value of $\lambda$,
the bound from Higgs stability rapidly becomes more stringent as the dimension
of the fermion representation increases.  With $G_x = SU(3)$, $r = \mathbf{3}$,
$\lambda = 1$, and $\alpha_x = 10\alpha$, the Higgs stability cutoff approaches
$\Lambda_H \simeq 3\,\tev$, only slightly above the range of explicit fermion
masses we are considering.  This situation can be improved somewhat by
lowering the new Yukawa coupling modestly; reducing to $\lambda = 0.8$
increases the stability cutoff scale to well over $10\,\tev$.  The corresponding
reduction in the Higgs branching fraction to dark vectors can be compensated
by the color factors in Eq.~\eqref{eq:hbrglu} and an increased
dark gauge coupling.  Note as well that in theories with non-Abelian 
dark ($SU(3)$) gauge groups motivated by electroweak naturalness, 
new physics is typically expected at scales below about 
$10\,\tev$~\cite{Burdman:2006tz,Cai:2008au,Craig:2015pha}.

  Direct collider searches for the massive fermions in the theory can be modified
in more radical ways by an unbroken non-Abelian dark gauge 
group with $\Lambda_x \ll m_1,\,m_2$~\cite{Okun:1980kw,Okun:1980mu,Kang:2008ea}.
Even so, we argue that our previous collider limits derived for the Abelian 
scenario can be applied here in many cases up to a rescaling by the 
fermion multiplicity $d(r)$.  The first stages of fermion production 
and decay proceed much like in the Abelian case.  Strong $G_x$ dynamics 
does not have a significant effect on fermion production (away from threshold), 
with the fermions created in pairs primarily by Drell-Yan processes.  
Next, the heavier $\psi_2$ and $P^-$ states decay down to the lightest 
$\psi_1$ mode.  For non-degenerate fermion masses, this typically occurs 
before the non-perturbative $G_x$ dynamics sets in.  

The immediate remnants of fermion production and electroweak cascade decays 
are therefore a $\psi_1\overline{\psi}_1$ pair. 
In contrast to the Abelian theory where they would leave the detector 
as missing energy, the fermions are now \emph{quirks} and remain
bound by a string of $G_x$ flux~\cite{Okun:1980kw,Okun:1980mu,Kang:2008ea}.  
This string eventually pulls the fermions back together, causing
them to oscillate until they annihilate~\cite{Kang:2008ea}.  
Since the $\psi_1$ fermions do not carry colour or electromagnetic charge, 
they do not interact significantly with the material in collider detectors
and they are not expected to be trapped.  
Their eventual annihilation products are dark glueballs, SM fermions, 
Higgs bosons, and $W$ and $Z$ vector bosons~\cite{Kang:2008ea,Martin:2010kk,Harnik:2011mv,Cheung:2008ke,Barger:1987xg,Chacko:2015fbc}.  
For $\Lambda_x\gtrsim 1\,\mev$, the dark glueball final states are the
dominant decay products.

  Our previous collider limits on the new fermions can be applied to
the non-Abelian scenario as well when the dark glueballs are the dominant
annihilation product and are long-lived.  When these two conditions are met,
the production modes and visible decay products are the nearly same as in 
the Abelian case up to an increased fermion multiplicity factor of $d(r)$.
For $\Lambda_x \lesssim 1\,\mev$, visible annihilation final states of the 
$\psi_1\overline{\psi}_1$ pair would provide an additional 
search channel~\cite{Cheung:2008ke,Chacko:2015fbc}.  With larger 
$\Lambda_x\gtrsim 1\,\gev$, displaced decays of the dark glueballs could 
be visible~\cite{Chacko:2015fbc}.

\subsection{Dark Matter Considerations}

  Thermal freezeout of $\psi_1$ proceeds similarly to the Abelian case, 
and can be treated in perturbation theory for $\Lambda_x \ll m_1$.
If the annihilation is dominated by $\psi\overline{\psi}_1 \to XX$ processes, 
the relic yield after freezeout is approximately
\beq
m_1Y_1 ~=~ m_1\lrf{n_1}{s} 
~\sim~ (10^{-11}\,\gev)\,\lrf{m_1}{500\,\gev}^2
\lrf{10\alpha}{\alpha_x}^2 \ .
\label{eq:yield}
\eeq
As the early universe cools to below $T \lesssim \Lambda_x$ after freezeout, 
the dark flux connections among the relic $\psi_1$ and $\overline{\psi}_1$
states become important.  These induce a second stage of 
$\psi_1\overline{\psi}_1$ annihilation to dark glueballs 
and SM final states~\cite{Kang:2008ea,Jacoby:2007nw,Nussinov:2009hc}.  

 For $\Lambda_x \gtrsim 100\,\mev$, this secondary annhilation is expected to 
occur before the onset of primordial nucleosynthesis~(BBN).
However, smaller $\Lambda_x$ produces a later stage of secondary 
annihilation, and the annihilation products can disrupt light
element abundances~\cite{Kawasaki:2004qu,Jedamzik:2006xz} 
or the cosmic microwave background radiation~\cite{Chen:2003gz,Fradette:2014sza}.
 We defer a full study of these effects to a future
work, but we note that these considerations suggest that larger values
of $\Lambda_x \gtrsim 100\,\mev$ are preferred if the new fermions and
glueballs were ever thermalized in the early universe.\footnote{  
Related considerations of relic glueball decays also tend to prefer larger
$\Lambda_x$ 
values~\cite{Boddy:2014yra,Boddy:2014qxa,Garcia:2015loa,Forestell:2016qhc}.}  
Note that these constraints differ significantly from theories with quirks 
that carry QCD color, in which a second stage of QCD annihilation reduces 
the quirk relic densities to acceptable 
levels~\cite{Kang:2008ea,Jacoby:2007nw,Nussinov:2009hc}.

  Cosmological constraints on relic fermions and glueballs can
also be avoided if their pre-decay relic yield is significantly below 
the thermal estimate of Eq.~\eqref{eq:yield}.  This can occur in scenarios
with low reheating temperatures~\cite{Gelmini:2006pw}, or even from the 
heavy $\psi_1$ fermions themselves if they come to dominate the energy
density of the universe before they decay~\cite{Soni:2017nlm}.

\section{Conclusions\label{sec:conc}}

 In this work we have studied the phenomenological consequences of 
the vectorized lepton portal, consisting of two or more new fermions
that are charged under both the SM and a dark gauge force and that 
connect to the Higgs boson through a Yukawa coupling.  The minimal
realization consists of electroweak singlet and doublet fermions and an
Abelian $U(1)_x$ dark gauge group.  These fermions act as mediators
between the visible and dark sectors, and they induce a gauge kinetic
mixing of the $U(1)_x$ vector with hypercharge.  

  An important consequence of the mediator fermions is new exotic
decay channels of the Higgs boson.  In particular, fermion loops induce 
$h\to XX$ and $h\to XZ$ decays.  The decay fractions of these modes
are potentially observable at the LHC for larger values of the new
Yukawa coupling $\lambda$ and the dark gauge coupling $\alpha_x$.
We find that existing LHC searches for $h\to XX$ constrain the product 
of the new neutral fermion masses to be at least 
$\sqrt{m_1m_2} \gtrsim 850\,\gev$ for $\lambda=1$, $\alpha_x = 10\alpha$, 
and $m_x= 15\,\gev$.  This sensitivity to exotic Higgs decays can be
significantly greater than direct limits on the new fermions from
precision electroweak tests, collider searches, and Higgs stability 
considerations.  Dark matter searches further constrain the new fermions, 
but the bounds depend on the evolution history of the cosmos.
As a result, searches for exotic Higgs decays with future data from 
the LHC and beyond are a key discovery channel for scenarios of this type.

  The minimal vectorized lepton portal studied here can also be extended
in a number of ways.  Expanding the new Yukawa coupling to a more general
chiral form allows for $CP$ violation in Higgs decays to dark 
vectors~\cite{Davoudiasl:2012ig,Voloshin:2012tv}.  If the dark sector
has spontaneous symmetry breaking, the new fermions can potentially mix
with SM leptons, leading to the violation of (charged) lepton flavor and
introducing new interactios among neutrinos.  The Abelian dark gauge group
we have concentrated on can also be extended to non-Abelian groups with
interesting consequences.  



\section*{Acknowledgements}

We thank Lindsay Forestell, David B. Kaplan, John Ng, Jessie Shelton, 
James Wells, and Richard Woloshyn for helpful discussions.
This work is supported by the Natural Sciences
and Engineering Research Council of Canada~(NSERC), with DM
supported in part by a Discovery Grant. 
TRIUMF receives federal funding via a contribution agreement 
with the National Research Council of Canada.

\appendix

\section{Higgs Loop Functions\label{app:hloop}}

\begin{figure}[ttt]
 \begin{center}
         \includegraphics[width = 0.5\textwidth]{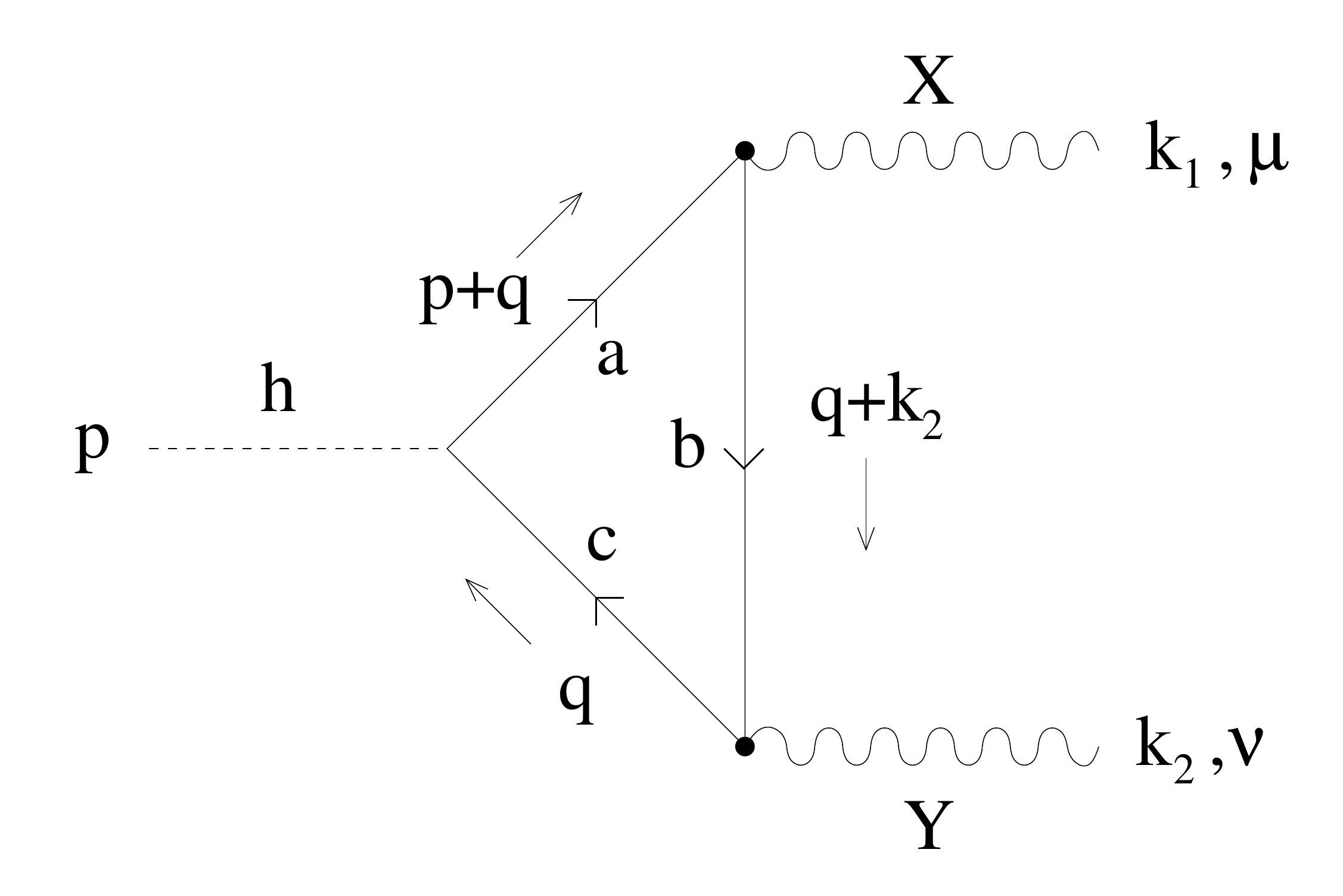}
 \end{center}
 \caption{Loop diagram for $h\to XY$ decay due to the fermion
loop $\{abc\}$.}
 \label{fig:hxy}
 \end{figure}

  Higgs boson decays to $h\to XX$ and $h\to XZ$ are generated
by loops of $\psi_1$ and $\psi_2$ fermions.  All the relevant
one-loop diagrams take the general form shown in Fig.~\ref{fig:hxy},
which connects the SM Higgs to a pair of vectors $X$ and $Y$  
with internal fermion states $a$, $b$, and $c$.  
It corresponds to a contribution to the amplitude of
\beq
-i\,(\Delta\mathcal{M}) = g_{ach}g_{cbY}g_{baX}\,
I^{\mu\nu}\varepsilon_{\mu}^*(k_1,\,\lambda_1)\varepsilon_{\nu}^*(k_2,\,\lambda_2)
\eeq
with
\beq
I^{\mu\nu} = -\int\!\frac{d^dq}{(2\pi)^d}\,
\frac{tr\left[(\fsl{q}+m_c)\gamma^{\nu}(\fsl{q}+\fsl{k}_1+m_b)\gamma^{\mu}
(\fsl{q}+\fsl{p}+m_a)\right]}
{(q^2-m_c^2+i\varepsilon)[(q+k_1-m_b)^2+i\varepsilon)][(q+p-m_a)^2+i\varepsilon)]} \ .
\eeq
For each such diagram, there is a second independent diagram
with the fermion arrows in Fig.~\ref{fig:hxy} reversed of the form 
\beq
I\!I^{\mu\nu} = 
I^{\mu\nu}(a\leftrightarrow c,\,k_1\leftrightarrow k_2,\mu\leftrightarrow \nu) 
\ .
\eeq

Computing the diagram with dimensional regularization in $d=(4-\epsilon)$, 
we find 
\beq
I^{\mu\nu} &=& \frac{4\,i}{(4\pi)^2}\,\int_0^1\!dx\int_0^{1-x}\!\!dy\;
\Bigg(\frac{1}{\Delta}
(m_am_bm_c\,\eta^{\mu\nu}+m_a\,A^{\mu\nu}+m_b\,B^{\mu\nu}+m_c\,C^{\mu\nu})
\label{eq:iint}\\
&&\hspace{3cm}-\eta^{\mu\nu}\left[(m_a+m_c-m_b) 
+ (2m_b-m_a-m_c)\left(\frac{2}{\epsilon} + \ldots\right)\right]
\Bigg) \ ,
\nnmb
\eeq
where
\beq
A^{\mu\nu} &=& 
\Big[-x^2\,m_h^2 + (x+y-2xy-y^2)\,m_Y^2 + (x-2xy)\,(k_1\ccdot k_2)\Big]\eta^{\mu\nu}
\label{eq:amn}\\
&& ~+~(-x+2x^2+2xy)\,k_1^{\nu}k_2^{\mu}
\nnmb\\
B^{\mu\nu} &=& 
\Big[(-x+x^2)\,m_h^2+(-y+2xy+y^2)\,m_Y^2+(-y+2xy)\,(k_1\ccdot k_2)\Big]\eta^{\mu\nu}
\label{eq:bmn}\\
&&~+~y\,k_1^{\nu}k_2^{\mu}
\nnmb\\
C^{\mu\nu} &=& \Big[(x-x^2)\,m_h^2 + (-1+x+2y-2xy-y^2)\,m_Y^2
\label{eq:cmn}\\
&&~~+~(-1+x+y-2xy)\,(k_1\ccdot k_2)\Big]\eta^{\mu\nu}
~~+~(1-3x-y + 2x^2+2xy)\,k_1^{\nu}k_2^{\mu}
\nnmb
\eeq
as well as
\beq
\Delta ~~=~~\Delta_{abc} &=& x\,m_a^2 + y\,m_b^2 + z\,m_c^2
\label{eq:delta}\\
&&~~+~(-x+x^2)\,m_h^2 + (-y+2xy+y^2)\,m_Y^2 
+ 2xy\,(k_1\ccdot k_2) \ . \nnmb
\eeq
The second loop $I\!I^{\mu\nu}$ can be obtained from this result 
by exchanging $a\leftrightarrow c$ everywhere.

  For $h\to XX$, the loops are $\{abc\} = \{111\},\,\{222\}$.
With $m_a=m_b=m_c$, we have $I^{\mu\nu}(aaa) = I\!I^{\mu\nu}(aaa)$
and the would-be divergent parts in Eq.~\eqref{eq:iint} cancel independently
in each.  The relevant coupling products are
\beq
g_{11h}\,g_{11X}\,g_{11X} &=&  
(-\sqrt{2}\lambda\,s_{\alpha}c_{\alpha})g_x^2
\\
g_{22h}\,g_{22X}\,g_{22X} &=& 
(+\sqrt{2}\lambda\,s_{\alpha}c_{\alpha})g_x^2 \ .
\eeq
Note the relative sign. 

  In the case of $h\to XZ$, the loops are
$\{abc\} = \{111\},\,\{222\},\,\{112\},\,\{221\}$, 
and the relevant coupling products are
\beq
g_{11h}\,g_{11X}\,g_{11Z} &=&  
(-\sqrt{2}\lambda\,s_{\alpha}c_{\alpha})(g_xq_x)(\bar{g}s_{\alpha}^2/2)
\\
g_{22h}\,g_{22X}\,g_{22Z} &=& 
(+\sqrt{2}\lambda\,s_{\alpha}c_{\alpha})(g_xq_x)(\bar{g}c_{\alpha}^2/2)
\\
g_{21h}\,g_{11X}\,g_{12Z} &=& [\lambda(c_{\alpha}^2-s_{\alpha}^2)/\sqrt{2}]
(g_xq_x)(-\bar{g}s_{\alpha}c_{\alpha}/2)
\\
g_{12h}\,g_{22X}\,g_{21Z} &=& g_{21h}\,g_{11X}\,g_{12Z}
\ ,
\eeq
where $\bar{g} = \sqrt{g^2+{g^\prime}^2}$.
For $\{111\}$ and $\{222\}$, the would-be divergent terms 
in Eq.~\eqref{eq:iint} cancel independently, 
while for $\{112\}$ and $\{221\}$ they cancel 
when the two contributions to the amplitude are summed.

\section{Electroweak Self-Energies\label{app:pew}}

The relevant loop functions in $d=(4-\epsilon)$ dimensions are
\beq
4\pi^2L_{ab}(p^2) &=& 
\left[\frac{1}{2}(m_a-m_b)^2-\frac{1}{3}p^2\right]
\left[\frac{2}{{\epsilon}}-\gamma_E+\ln(4\pi) - \ln\lrf{p^2}{\mu^2}\right]\\
&&~~~+ \left[(m_am_b-m_a^2)\,\widetilde{b}_0 +(m_a^2-m_b^2+2p^2)\,\widetilde{b}_1
-2p^2\,\widetilde{b}_2\right] \ ,
\nnmb
\eeq
where $\mu$ is the renormalization scale and 
\beq
\widetilde{b}_0(p,m_a,m_b) &=& 
\sum_{i=\pm}\left[\ln(1-x_i)-x_i\ln\left(1-\frac{1}{x_i}\right) - 1\right]
\\
2\,\widetilde{b}_1(p,m_a,m_b) &=& 
\sum_{i=\pm}\left[\ln(1-x_i)-x_i^2\ln\left(1-\frac{1}{x_i}\right) -x_i - \frac{1}{2}\right]
\\
3\,\widetilde{b}_2(p,m_a,m_b) &=& 
\sum_{i=\pm}\left[\ln(1-x_i)-x_i^3\ln\left(1-\frac{1}{x_i}\right) 
- x_i^2 - \frac{x_i}{2} - \frac{1}{3}\right]
\eeq
in which the index $i$ labels
\beq
x_{\pm} = \frac{1}{2p^2}\left[(p^2+m_a^2-m_b^2)\pm
\sqrt{(p^2+m_a^2-m_b^2)^2-4p^2(m_a^2-i\varepsilon)}\right] \ ,
\eeq
and the $i\varepsilon$ defines the proper branch of the logarithms
when their arguments become negative or complex. These loop functions 
are closely related to (the finite parts) of the Passarino-Veltman
functions~\cite{Passarino:1978jh}.
For $p^2\to 0$, the result simplifies to
\beq
4\pi^2L_{ab}(p^2) &=&
\frac{1}{2}(m_a-m_b)^2
\left[\frac{2}{{\epsilon}}-\gamma_E+\ln(4\pi) - \ln\lrf{m_am_b}{\mu^2}
+\frac{1}{2}
\right]\\
&&~
-\frac{1}{2}m_am_b
-\frac{1}{4(m_a^2-m_b^2)}\ln\lrf{m_a^2}{m_b^2}
\left(m_a^4-2m_a^3{m}_b - 2m_am_b^3 + m_b^4\right) \ .
\nnmb
\eeq

In terms of these loop functions, the shifts in the oblique
parameters $S$, $T$, and $U$ due to the vector-like fermions 
are~\cite{Peskin:1990zt,Peskin:1991sw}
\beq
\Delta S &=& \frac{4\pi}{m_Z^2}\left(
-\left[L_{--}(m_Z^2)-L_{--}(0)\right]\phantom{\frac{I}{I}}
\right.\\
&&
~~~~~\left.\phantom{\frac{I}{I}}
+ s_{\alpha}^4\left[L_{11}(m_Z^2)-L_{11}(0)\right]
+2c_{\alpha}^2s_{\alpha}^2\left[12\right]+c_{\alpha}^4[22]~\right)
\nnmb\\
\Delta S+\Delta U &=& \frac{8\pi}{c_W^2m_Z^2}\left(\phantom{\frac{I}{I}}\!\!
s_{\alpha}^2\left[L_{1-}(m_W^2)-L_{1-}(0)\right]
+c_{\alpha}^2\left[L_{2-}(m_W^2)-L_{2-}(0)\right]
\right.
\\
&&
\left.
~~~~~~~~
- {c_W^2}\left[L_{--}(m_Z^2)-L_{--}(0)\right]
\!\!\phantom{\frac{I}{I}}
\right) 
\nnmb\\
\Delta T &=& \frac{2\pi}{s_W^2c_W^2\,m_Z^2}\left[\phantom{\!\!\frac{I}{I}}
s_{\alpha}^2L_{1-}(0) + c_{\alpha}^2L_{2-}(0)
- s_{\alpha}^2c_{\alpha}^2L_{12}(0)\phantom{\frac{I}{I}}\!\!\right] \ .
\eeq
These expressions are independent of $1/{\epsilon}$ and the 
renormalization scale $\mu$.

\section{Renormalization Group Equations\label{app:rg}}

  We collect here the one-loop renormalization group~(RG) equations
relevant for the Higgs stability analysis of Sec.~\ref{sec:stability}.
In these equations, the only the SM Yukawa coupling we keep is that
of the top quark, and we use the $SU(5)$ normalization for the
hypercharge coupling, $g_1 = \sqrt{5/3}\,g^\prime$.
Our normalization for the Higgs self coupling is
$V(H) \supset \lambda_H|H|^4$ so that $\lambda_H \simeq m_h^2/2v^2$
with $v\simeq 174\,\gev$.  To allow for generalization beyond
the minimal Abelian vectorized lepton portal theory, we write the RG
equations for general dark gauge group $G_x$ under which
$P$ and $N$ transform under the representation $r_x$
with dimension $d(r_x)$.  

  With these assumptions, the RG equations for the system 
above the heavy fermion threshold can be adapted from
the general results of 
Refs.~\cite{Machacek:1983tz,Machacek:1983fi,Machacek:1984zw}
as in Refs.~\cite{Kribs:2007nz,Hashimoto:2010at,Ishiwata:2011hr,ArkaniHamed:2012kq}.
We find
\beq
(4\pi)^2\frac{d\lambda_H}{dt} &=& 24\,\lambda_H^2
+ 4\lambda_H[3y_t^2+2d(r_x)\lambda^2] - 2[3y_t^4+2d(r_x)\lambda^4]\\
&&
- 3\lambda_H(3g_2^2+\frac{3}{5}g_1^2)
+ \frac{3}{8}\left[2g_2^4+(g_2^2+\frac{3}{5}g_1^2)^2\right]
\phantom{\frac{\hat{I}}{\hat{I}}}
\nnmb\\
(4\pi)^2\frac{dy_t}{dt} &=& \frac{9}{2}y_t^3 + 2d(r_x)y_t\lambda^2 
-y_t(8g_3^2+\frac{9}{4}g_2^2+\frac{17}{20}g_1^2) 
\phantom{\frac{\hat{I}}{\hat{I}}}\\
(4\pi)^2\frac{d\lambda}{dt} &=& \left[\frac{3+4d(r_x)}{2}\right]\lambda^3
+ 3\lambda y_t^2  
- \lambda[\frac{9}{4}g_2^2+\frac{9}{20}g_1^2+6C_2(r_x)g_x^2]
\phantom{\frac{\hat{I}}{\hat{I}}} \ ,
\eeq
together with
\beq
(4\pi)^2\frac{dg_2}{dt} &=& \left[-\frac{19}{6}+\frac{2}{3}d(r_x)\right]g_2^3\\
(4\pi)^2\frac{dg_1}{dt} &=& \left[\frac{41}{10}+\frac{2}{5}d(r_x)\right]g_1^3\\
(4\pi)^2\frac{dg_x}{dt} &=& \left[-\frac{11}{3}C_2(G_x) + 4S_2(r_x)\right]g_x^3
 \ ,
\eeq
where $t=\ln(\mu/\mu_0)$ defines the renormalization scale,
and $S_2(r_x)$ and $C_2(r_x)$ refer to the trace and Casimir invariants
of the representation $r_x$ of $N$ and $P$ under $G_x$.

For $G_x = U(1)_x$ with $N_f$ copies of the $N$ and $P$ fields 
of charge $q_x$, we have
\beq
C_2(r_x) = q_x^2 \ ,~~~~~S_2(r_x) = q_x^2N_f \ ,~~~~~d(r_x) = N_f \ ,~~~~~
C_2(G_x) = 0 \ .
\eeq
This case also allows for kinetic mixing between hypercharge and $U(1)_x$.
The corresponding evolution equation for the mixing parameter
$\tilde{\epsilon} = \epsilon/c_W$ above the heavy fermion mass threshold 
is~\cite{delAguila:1988jz,Dienes:1996zr}
\beq
(4\pi)^2\frac{d\tilde\epsilon}{dt} &=& 4N_f\tilde\epsilon\,(g_xq_x)^2
+ \large(\frac{41}{10}+\frac{2}{5}N_f\large)\tilde\epsilon\,g_1^2
- \frac{8}{3}\sqrt{\frac{3}{5}}\,N_f\,g_1(g_xq_x) \ .
\eeq
Below the heavy fermion masses, the remaining evolution is 
homogeneous in $\tilde\epsilon$.  There is also a small correction
to the running of $g_1$ and $g_x$ proportional to $\tilde\epsilon$
that we do not include.


\bibliographystyle{unsrt}

\begin{thebibliography}{9}

\bibitem{Fayet:2007ua} 
  P.~Fayet,
  Phys.\ Rev.\ D {\bf 75}, 115017 (2007)
  doi:10.1103/PhysRevD.75.115017
  [hep-ph/0702176 [HEP-PH]].

\bibitem{Pospelov:2008zw} 
  M.~Pospelov,
  Phys.\ Rev.\ D {\bf 80}, 095002 (2009)
  doi:10.1103/PhysRevD.80.095002
  [arXiv:0811.1030 [hep-ph]].

\bibitem{Bjorken:2009mm} 
  J.~D.~Bjorken, R.~Essig, P.~Schuster and N.~Toro,
  Phys.\ Rev.\ D {\bf 80}, 075018 (2009)
  doi:10.1103/PhysRevD.80.075018
  [arXiv:0906.0580 [hep-ph]].

\bibitem{Essig:2013lka} 
  R.~Essig {\it et al.},
  arXiv:1311.0029 [hep-ph].

\bibitem{Alexander:2016aln} 
  J.~Alexander {\it et al.},
  arXiv:1608.08632 [hep-ph].

\bibitem{Boehm:2003hm} 
  C.~Boehm and P.~Fayet,
  Nucl.\ Phys.\ B {\bf 683}, 219 (2004)
  doi:10.1016/j.nuclphysb.2004.01.015
  [hep-ph/0305261].

\bibitem{Borodatchenkova:2005ct} 
  N.~Borodatchenkova, D.~Choudhury and M.~Drees,
  Phys.\ Rev.\ Lett.\  {\bf 96}, 141802 (2006)
  doi:10.1103/PhysRevLett.96.141802
  [hep-ph/0510147].

\bibitem{Pospelov:2007mp} 
  M.~Pospelov, A.~Ritz and M.~B.~Voloshin,
  Phys.\ Lett.\ B {\bf 662}, 53 (2008)
  doi:10.1016/j.physletb.2008.02.052
  [arXiv:0711.4866 [hep-ph]].

\bibitem{ArkaniHamed:2008qn} 
  N.~Arkani-Hamed, D.~P.~Finkbeiner, T.~R.~Slatyer and N.~Weiner,
  Phys.\ Rev.\ D {\bf 79}, 015014 (2009)
  doi:10.1103/PhysRevD.79.015014
  [arXiv:0810.0713 [hep-ph]].

\bibitem{Okun:1982xi} 
  L.~B.~Okun,
  Sov.\ Phys.\ JETP {\bf 56}, 502 (1982)
  [Zh.\ Eksp.\ Teor.\ Fiz.\  {\bf 83}, 892 (1982)].

\bibitem{Holdom:1985ag} 
  B.~Holdom,
  Phys.\ Lett.\ B {\bf 166}, 196 (1986).
  doi:10.1016/0370-2693(86)91377-8

\bibitem{Schabinger:2005ei} 
  R.~M.~Schabinger and J.~D.~Wells,
  Phys.\ Rev.\ D {\bf 72}, 093007 (2005)
  doi:10.1103/PhysRevD.72.093007
  [hep-ph/0509209].

\bibitem{Patt:2006fw} 
  B.~Patt and F.~Wilczek,
  hep-ph/0605188.

\bibitem{Strassler:2006im} 
  M.~J.~Strassler and K.~M.~Zurek,
  Phys.\ Lett.\ B {\bf 651}, 374 (2007)
  doi:10.1016/j.physletb.2007.06.055
  [hep-ph/0604261].

\bibitem{Strassler:2006qa} 
  M.~J.~Strassler,
  hep-ph/0607160.

\bibitem{Han:2007ae} 
  T.~Han, Z.~Si, K.~M.~Zurek and M.~J.~Strassler,
  JHEP {\bf 0807}, 008 (2008)
  doi:10.1088/1126-6708/2008/07/008
  [arXiv:0712.2041 [hep-ph]].

\bibitem{Davoudiasl:2012ig} 
  H.~Davoudiasl, H.~S.~Lee and W.~J.~Marciano,
  Phys.\ Rev.\ D {\bf 86}, 095009 (2012)
  doi:10.1103/PhysRevD.86.095009
  [arXiv:1208.2973 [hep-ph]].


\bibitem{Stueckelberg:1900zz} 
  E.~C.~G.~Stueckelberg,
  Helv.\ Phys.\ Acta {\bf 11}, 225 (1938).
  doi:10.5169/seals-110852

\bibitem{Kors:2004dx} 
  B.~Kors and P.~Nath,
  Phys.\ Lett.\ B {\bf 586}, 366 (2004)
  doi:10.1016/j.physletb.2004.02.051
  [hep-ph/0402047].

\bibitem{Ellis:2014dza} 
  S.~A.~R.~Ellis, R.~M.~Godbole, S.~Gopalakrishna and J.~D.~Wells,
  JHEP {\bf 1409}, 130 (2014)
  doi:10.1007/JHEP09(2014)130
  [arXiv:1404.4398 [hep-ph]].

\bibitem{Bhattacharya:2015qpa} 
  S.~Bhattacharya, N.~Sahoo and N.~Sahu,
  Phys.\ Rev.\ D {\bf 93}, no. 11, 115040 (2016)
  doi:10.1103/PhysRevD.93.115040
  [arXiv:1510.02760 [hep-ph]].

\bibitem{Dvali:1996cu} 
  G.~R.~Dvali, G.~F.~Giudice and A.~Pomarol,
  Nucl.\ Phys.\ B {\bf 478}, 31 (1996)
  doi:10.1016/0550-3213(96)00404-X
  [hep-ph/9603238].

\bibitem{Craig:2012xp} 
  N.~Craig, S.~Knapen, D.~Shih and Y.~Zhao,
  JHEP {\bf 1303}, 154 (2013)
  doi:10.1007/JHEP03(2013)154
  [arXiv:1206.4086 [hep-ph]].

\bibitem{Azatov:2011ht} 
  A.~Azatov, J.~Galloway and M.~A.~Luty,
  Phys.\ Rev.\ Lett.\  {\bf 108}, 041802 (2012)
  doi:10.1103/PhysRevLett.108.041802
  [arXiv:1106.3346 [hep-ph]].

\bibitem{Heckman:2011bb} 
  J.~J.~Heckman, P.~Kumar, C.~Vafa and B.~Wecht,
  JHEP {\bf 1201}, 156 (2012)
  doi:10.1007/JHEP01(2012)156
  [arXiv:1108.3849 [hep-ph]].

\bibitem{Evans:2012uf} 
  J.~L.~Evans, M.~Ibe and T.~T.~Yanagida,
  Phys.\ Rev.\ D {\bf 86}, 015017 (2012)
  doi:10.1103/PhysRevD.86.015017
  [arXiv:1204.6085 [hep-ph]].

\bibitem{Burdman:2006tz} 
  G.~Burdman, Z.~Chacko, H.~S.~Goh and R.~Harnik,
  JHEP {\bf 0702}, 009 (2007)
  doi:10.1088/1126-6708/2007/02/009
  [hep-ph/0609152].

\bibitem{Cai:2008au} 
  H.~Cai, H.~C.~Cheng and J.~Terning,
  JHEP {\bf 0905}, 045 (2009)
  doi:10.1088/1126-6708/2009/05/045
  [arXiv:0812.0843 [hep-ph]].

\bibitem{Graham:2015cka} 
  P.~W.~Graham, D.~E.~Kaplan and S.~Rajendran,
  Phys.\ Rev.\ Lett.\  {\bf 115}, no. 22, 221801 (2015)
  doi:10.1103/PhysRevLett.115.221801
  [arXiv:1504.07551 [hep-ph]].

\bibitem{Antipin:2015jia} 
  O.~Antipin and M.~Redi,
  JHEP {\bf 1512}, 031 (2015)
  doi:10.1007/JHEP12(2015)031
  [arXiv:1508.01112 [hep-ph]].

\bibitem{Batell:2015fma} 
  B.~Batell, G.~F.~Giudice and M.~McCullough,
  JHEP {\bf 1512}, 162 (2015)
  doi:10.1007/JHEP12(2015)162
  [arXiv:1509.00834 [hep-ph]].

\bibitem{Choi:2015fiu} 
  K.~Choi and S.~H.~Im,
  JHEP {\bf 1601}, 149 (2016)
  doi:10.1007/JHEP01(2016)149
  [arXiv:1511.00132 [hep-ph]].

\bibitem{Davoudiasl:2013aya} 
  H.~Davoudiasl, H.~S.~Lee, I.~Lewis and W.~J.~Marciano,
  Phys.\ Rev.\ D {\bf 88}, no. 1, 015022 (2013)
  doi:10.1103/PhysRevD.88.015022
  [arXiv:1304.4935 [hep-ph]].

\bibitem{DiFranzo:2015nli} 
  A.~DiFranzo, P.~J.~Fox and T.~M.~P.~Tait,
  JHEP {\bf 1604}, 135 (2016)
  doi:10.1007/JHEP04(2016)135
  [arXiv:1512.06853 [hep-ph]].

\bibitem{DiFranzo:2016uzc} 
  A.~DiFranzo and G.~Mohlabeng,
  JHEP {\bf 1701}, 080 (2017)
  doi:10.1007/JHEP01(2017)080
  [arXiv:1610.07606 [hep-ph]].

\bibitem{Juknevich:2009gg} 
  J.~E.~Juknevich,
  JHEP {\bf 1008}, 121 (2010)
  doi:10.1007/JHEP08(2010)121
  [arXiv:0911.5616 [hep-ph]].

\bibitem{Beauchesne:2017ukw} 
  H.~Beauchesne, E.~Bertuzzo and G.~Grilli di Cortona,
  arXiv:1705.06325 [hep-ph].

\bibitem{Mahbubani:2005pt} 
  R.~Mahbubani and L.~Senatore,
  Phys.\ Rev.\ D {\bf 73}, 043510 (2006)
  doi:10.1103/PhysRevD.73.043510
  [hep-ph/0510064].

\bibitem{DEramo:2007anh} 
  F.~D'Eramo,
  Phys.\ Rev.\ D {\bf 76}, 083522 (2007)
  doi:10.1103/PhysRevD.76.083522
  [arXiv:0705.4493 [hep-ph]].

\bibitem{Essig:2007az} 
  R.~Essig,
  Phys.\ Rev.\ D {\bf 78}, 015004 (2008)
  doi:10.1103/PhysRevD.78.015004
  [arXiv:0710.1668 [hep-ph]].

\bibitem{Cohen:2011ec} 
  T.~Cohen, J.~Kearney, A.~Pierce and D.~Tucker-Smith,
  Phys.\ Rev.\ D {\bf 85}, 075003 (2012)
  doi:10.1103/PhysRevD.85.075003
  [arXiv:1109.2604 [hep-ph]].

\bibitem{Davoudiasl:2012ag} 
  H.~Davoudiasl, H.~S.~Lee and W.~J.~Marciano,
  Phys.\ Rev.\ D {\bf 85}, 115019 (2012)
  doi:10.1103/PhysRevD.85.115019
  [arXiv:1203.2947 [hep-ph]].

\bibitem{delAguila:1988jz} 
  F.~del Aguila, G.~D.~Coughlan and M.~Quiros,
  Nucl.\ Phys.\ B {\bf 307}, 633 (1988)
  Erratum: [Nucl.\ Phys.\ B {\bf 312}, 751 (1989)].
  doi:10.1016/0550-3213(88)90266-0

\bibitem{Dienes:1996zr} 
  K.~R.~Dienes, C.~F.~Kolda and J.~March-Russell,
  Nucl.\ Phys.\ B {\bf 492}, 104 (1997)
  doi:10.1016/S0550-3213(97)80028-4, 10.1016/S0550-3213(97)00173-9
  [hep-ph/9610479].

\bibitem{Shifman:1979eb} 
  M.~A.~Shifman, A.~I.~Vainshtein, M.~B.~Voloshin and V.~I.~Zakharov,
  Sov.\ J.\ Nucl.\ Phys.\  {\bf 30}, 711 (1979)
  [Yad.\ Fiz.\  {\bf 30}, 1368 (1979)].

\bibitem{Aad:2015sva} 
  G.~Aad {\it et al.} [ATLAS Collaboration],
  Phys.\ Rev.\ D {\bf 92}, no. 9, 092001 (2015)
  doi:10.1103/PhysRevD.92.092001
  [arXiv:1505.07645 [hep-ex]].

\bibitem{Gopalakrishna:2008dv} 
  S.~Gopalakrishna, S.~Jung and J.~D.~Wells,
  Phys.\ Rev.\ D {\bf 78}, 055002 (2008)
  doi:10.1103/PhysRevD.78.055002
  [arXiv:0801.3456 [hep-ph]].

\bibitem{Curtin:2013fra} 
  D.~Curtin {\it et al.},
  Phys.\ Rev.\ D {\bf 90}, no. 7, 075004 (2014)
  doi:10.1103/PhysRevD.90.075004
  [arXiv:1312.4992 [hep-ph]].

\bibitem{Falkowski:2014ffa} 
  A.~Falkowski and R.~Vega-Morales,
  JHEP {\bf 1412}, 037 (2014)
  doi:10.1007/JHEP12(2014)037
  [arXiv:1405.1095 [hep-ph]].

\bibitem{Curtin:2014cca} 
  D.~Curtin, R.~Essig, S.~Gori and J.~Shelton,
  JHEP {\bf 1502}, 157 (2015)
  doi:10.1007/JHEP02(2015)157
  [arXiv:1412.0018 [hep-ph]].

\bibitem{Gabrielli:2014oya} 
  E.~Gabrielli, M.~Heikinheimo, B.~Mele and M.~Raidal,
  Phys.\ Rev.\ D {\bf 90}, no. 5, 055032 (2014)
  doi:10.1103/PhysRevD.90.055032
  [arXiv:1405.5196 [hep-ph]].

\bibitem{Biswas:2015sha} 
  S.~Biswas, E.~Gabrielli, M.~Heikinheimo and B.~Mele,
  JHEP {\bf 1506}, 102 (2015)
  doi:10.1007/JHEP06(2015)102
  [arXiv:1503.05836 [hep-ph]].

\bibitem{Biswas:2016jsh} 
  S.~Biswas, E.~Gabrielli, M.~Heikinheimo and B.~Mele,
  Phys.\ Rev.\ D {\bf 93}, no. 9, 093011 (2016)
  doi:10.1103/PhysRevD.93.093011
  [arXiv:1603.01377 [hep-ph]].

\bibitem{Biswas:2017lyg} 
  S.~Biswas, E.~Gabrielli, M.~Heikinheimo and B.~Mele,
  arXiv:1703.00402 [hep-ph].

\bibitem{Campos:2017dgc} 
  M.~D.~Campos, D.~Cogollo, M.~Lindner, T.~Melo, F.~S.~Queiroz and W.~Rodejohann,
  arXiv:1705.05388 [hep-ph].

\bibitem{Hook:2010tw} 
  A.~Hook, E.~Izaguirre and J.~G.~Wacker,
  Adv.\ High Energy Phys.\  {\bf 2011}, 859762 (2011)
  doi:10.1155/2011/859762
  [arXiv:1006.0973 [hep-ph]].

\bibitem{Falkowski:2010cm} 
  A.~Falkowski, J.~T.~Ruderman, T.~Volansky and J.~Zupan,
  JHEP {\bf 1005}, 077 (2010)
  doi:10.1007/JHEP05(2010)077
  [arXiv:1002.2952 [hep-ph]].

\bibitem{Falkowski:2010gv} 
  A.~Falkowski, J.~T.~Ruderman, T.~Volansky and J.~Zupan,
  Phys.\ Rev.\ Lett.\  {\bf 105}, 241801 (2010)
  doi:10.1103/PhysRevLett.105.241801
  [arXiv:1007.3496 [hep-ph]].

\bibitem{Peskin:1990zt} 
  M.~E.~Peskin and T.~Takeuchi,
  Phys.\ Rev.\ Lett.\  {\bf 65}, 964 (1990).
  doi:10.1103/PhysRevLett.65.964

\bibitem{Peskin:1991sw} 
  M.~E.~Peskin and T.~Takeuchi,
  Phys.\ Rev.\ D {\bf 46}, 381 (1992).
  doi:10.1103/PhysRevD.46.381

\bibitem{Babu:1997st} 
  K.~S.~Babu, C.~F.~Kolda and J.~March-Russell,
  Phys.\ Rev.\ D {\bf 57}, 6788 (1998)
  doi:10.1103/PhysRevD.57.6788
  [hep-ph/9710441].

\bibitem{Kumar:2006gm} 
  J.~Kumar and J.~D.~Wells,
  Phys.\ Rev.\ D {\bf 74}, 115017 (2006)
  doi:10.1103/PhysRevD.74.115017
  [hep-ph/0606183].

\bibitem{Chang:2006fp} 
  W.~F.~Chang, J.~N.~Ng and J.~M.~S.~Wu,
  Phys.\ Rev.\ D {\bf 74}, 095005 (2006)
  Erratum: [Phys.\ Rev.\ D {\bf 79}, 039902 (2009)]
  doi:10.1103/PhysRevD.74.095005, 10.1103/PhysRevD.79.039902
  [hep-ph/0608068].

\bibitem{Chen:2017hak} 
  C.~Y.~Chen, S.~Dawson and E.~Furlan,
  arXiv:1703.06134 [hep-ph].

\bibitem{Baak:2014ora} 
  M.~Baak {\it et al.} [Gfitter Group],
  Eur.\ Phys.\ J.\ C {\bf 74}, 3046 (2014)
  doi:10.1140/epjc/s10052-014-3046-5
  [arXiv:1407.3792 [hep-ph]].

\bibitem{Kumar:2015tna} 
  N.~Kumar and S.~P.~Martin,
  Phys.\ Rev.\ D {\bf 92}, no. 11, 115018 (2015)
  doi:10.1103/PhysRevD.92.115018
  [arXiv:1510.03456 [hep-ph]].

\bibitem{lepsusy}
LEPSUSYWG, ALEPH, DELPHI, L3 and OPAL experiments,\\
\href{https://lepsusy.web.cern.ch/lepsusy/}{http://lepsusy.web.cern.ch/lepsusy/}

\bibitem{lepchg1}
LEPSUSYWG, ALEPH, DELPHI, L3 and OPAL experiments,
note LEPSUSYWG/01-03.1\\
\href{https://lepsusy.web.cern.ch/lepsusy/}{http://lepsusy.web.cern.ch/lepsusy/}

\bibitem{Fox:2011fx} 
  P.~J.~Fox, R.~Harnik, J.~Kopp and Y.~Tsai,
  Phys.\ Rev.\ D {\bf 84}, 014028 (2011)
  doi:10.1103/PhysRevD.84.014028
  [arXiv:1103.0240 [hep-ph]].

\bibitem{Abdallah:2003np} 
  J.~Abdallah {\it et al.} [DELPHI Collaboration],
  Eur.\ Phys.\ J.\ C {\bf 38}, 395 (2005)
  doi:10.1140/epjc/s2004-02051-8
  [hep-ex/0406019].

\bibitem{Buckley:2014fba} 
  M.~R.~Buckley, D.~Feld and D.~Goncalves,
  Phys.\ Rev.\ D {\bf 91}, 015017 (2015)
  doi:10.1103/PhysRevD.91.015017
  [arXiv:1410.6497 [hep-ph]].

\bibitem{Harris:2014hga} 
  P.~Harris, V.~V.~Khoze, M.~Spannowsky and C.~Williams,
  Phys.\ Rev.\ D {\bf 91}, 055009 (2015)
  doi:10.1103/PhysRevD.91.055009
  [arXiv:1411.0535 [hep-ph]].

\bibitem{Craig:2014lda} 
  N.~Craig, H.~K.~Lou, M.~McCullough and A.~Thalapillil,
  JHEP {\bf 1602}, 127 (2016)
  doi:10.1007/JHEP02(2016)127
  [arXiv:1412.0258 [hep-ph]].

\bibitem{CMS:2017vwf} 
  CMS Collaboration [CMS Collaboration],
  CMS-PAS-SUS-16-034.

\bibitem{Alwall:2014hca} 
  J.~Alwall {\it et al.},
  JHEP {\bf 1407}, 079 (2014)
  doi:10.1007/JHEP07(2014)079
  [arXiv:1405.0301 [hep-ph]].

\bibitem{Alloul:2013bka} 
  A.~Alloul, N.~D.~Christensen, C.~Degrande, C.~Duhr and B.~Fuks,
  Comput.\ Phys.\ Commun.\  {\bf 185}, 2250 (2014)
  doi:10.1016/j.cpc.2014.04.012
  [arXiv:1310.1921 [hep-ph]].

\bibitem{CMS:2017fdz} 
  CMS Collaboration [CMS Collaboration],
  CMS-PAS-SUS-16-039.

\bibitem{CMS:2017fij} 
  CMS Collaboration [CMS Collaboration],
  CMS-PAS-SUS-16-048.

\bibitem{Kribs:2007nz} 
  G.~D.~Kribs, T.~Plehn, M.~Spannowsky and T.~M.~P.~Tait,
  Phys.\ Rev.\ D {\bf 76}, 075016 (2007)
  doi:10.1103/PhysRevD.76.075016
  [arXiv:0706.3718 [hep-ph]].

\bibitem{Hashimoto:2010at} 
  M.~Hashimoto,
  Phys.\ Rev.\ D {\bf 81}, 075023 (2010)
  doi:10.1103/PhysRevD.81.075023
  [arXiv:1001.4335 [hep-ph]].

\bibitem{Ishiwata:2011hr} 
  K.~Ishiwata and M.~B.~Wise,
  Phys.\ Rev.\ D {\bf 84}, 055025 (2011)
  doi:10.1103/PhysRevD.84.055025
  [arXiv:1107.1490 [hep-ph]].

\bibitem{ArkaniHamed:2012kq} 
  N.~Arkani-Hamed, K.~Blum, R.~T.~D'Agnolo and J.~Fan,
  JHEP {\bf 1301}, 149 (2013)
  doi:10.1007/JHEP01(2013)149
  [arXiv:1207.4482 [hep-ph]].

\bibitem{Isidori:2001bm} 
  G.~Isidori, G.~Ridolfi and A.~Strumia,
  Nucl.\ Phys.\ B {\bf 609}, 387 (2001)
  doi:10.1016/S0550-3213(01)00302-9
  [hep-ph/0104016].

\bibitem{Buttazzo:2013uya} 
  D.~Buttazzo, G.~Degrassi, P.~P.~Giardino, G.~F.~Giudice, F.~Sala, A.~Salvio and A.~Strumia,
  JHEP {\bf 1312}, 089 (2013)
  doi:10.1007/JHEP12(2013)089
  [arXiv:1307.3536 [hep-ph]].

\bibitem{Machacek:1983tz} 
  M.~E.~Machacek and M.~T.~Vaughn,
  Nucl.\ Phys.\ B {\bf 222}, 83 (1983).
  doi:10.1016/0550-3213(83)90610-7

\bibitem{Machacek:1983fi} 
  M.~E.~Machacek and M.~T.~Vaughn,
  Nucl.\ Phys.\ B {\bf 236}, 221 (1984).
  doi:10.1016/0550-3213(84)90533-9

\bibitem{Machacek:1984zw} 
  M.~E.~Machacek and M.~T.~Vaughn,
  Nucl.\ Phys.\ B {\bf 249}, 70 (1985).
  doi:10.1016/0550-3213(85)90040-9

\bibitem{Martin:2016xsp} 
  S.~P.~Martin,
  Phys.\ Rev.\ D {\bf 93}, no. 9, 094017 (2016)
  doi:10.1103/PhysRevD.93.094017
  [arXiv:1604.01134 [hep-ph]].

\bibitem{Cline:2014dwa} 
  J.~M.~Cline, G.~Dupuis, Z.~Liu and W.~Xue,
  JHEP {\bf 1408}, 131 (2014)
  doi:10.1007/JHEP08(2014)131
  [arXiv:1405.7691 [hep-ph]].

\bibitem{Baer:1998pg} 
  H.~Baer, K.~m.~Cheung and J.~F.~Gunion,
  Phys.\ Rev.\ D {\bf 59}, 075002 (1999)
  doi:10.1103/PhysRevD.59.075002
  [hep-ph/9806361].

\bibitem{Lattanzi:2008qa} 
  M.~Lattanzi and J.~I.~Silk,
  Phys.\ Rev.\ D {\bf 79}, 083523 (2009)
  doi:10.1103/PhysRevD.79.083523
  [arXiv:0812.0360 [astro-ph]].

\bibitem{Falkowski:2009yz} 
  A.~Falkowski, J.~Juknevich and J.~Shelton,
  arXiv:0908.1790 [hep-ph].

\bibitem{Jungman:1995df} 
  G.~Jungman, M.~Kamionkowski and K.~Griest,
  Phys.\ Rept.\  {\bf 267}, 195 (1996)
  doi:10.1016/0370-1573(95)00058-5
  [hep-ph/9506380].

\bibitem{Halverson:2014nwa} 
  J.~Halverson, N.~Orlofsky and A.~Pierce,
  Phys.\ Rev.\ D {\bf 90}, no. 1, 015002 (2014)
  doi:10.1103/PhysRevD.90.015002
  [arXiv:1403.1592 [hep-ph]].

\bibitem{Hill:2014yxa} 
  R.~J.~Hill and M.~P.~Solon,
  Phys.\ Rev.\ D {\bf 91}, 043505 (2015)
  doi:10.1103/PhysRevD.91.043505
  [arXiv:1409.8290 [hep-ph]].

\bibitem{Bishara:2016hek} 
  F.~Bishara, J.~Brod, B.~Grinstein and J.~Zupan,
  JCAP {\bf 1702}, no. 02, 009 (2017)
  doi:10.1088/1475-7516/2017/02/009
  [arXiv:1611.00368 [hep-ph]].

\bibitem{Junnarkar:2013ac} 
  P.~Junnarkar and A.~Walker-Loud,
  Phys.\ Rev.\ D {\bf 87}, 114510 (2013)
  doi:10.1103/PhysRevD.87.114510
  [arXiv:1301.1114 [hep-lat]].

\bibitem{Tan:2016zwf} 
  A.~Tan {\it et al.} [PandaX-II Collaboration],
  Phys.\ Rev.\ Lett.\  {\bf 117}, no. 12, 121303 (2016)
  doi:10.1103/PhysRevLett.117.121303
  [arXiv:1607.07400 [hep-ex]].

\bibitem{Akerib:2016vxi} 
  D.~S.~Akerib {\it et al.} [LUX Collaboration],
  Phys.\ Rev.\ Lett.\  {\bf 118}, no. 2, 021303 (2017)
  doi:10.1103/PhysRevLett.118.021303
  [arXiv:1608.07648 [astro-ph.CO]].

\bibitem{Ahmed:2009zw} 
  Z.~Ahmed {\it et al.} [CDMS-II Collaboration],
  Science {\bf 327}, 1619 (2010)
  doi:10.1126/science.1186112
  [arXiv:0912.3592 [astro-ph.CO]].

\bibitem{Gelmini:2006pw} 
  G.~B.~Gelmini and P.~Gondolo,
  Phys.\ Rev.\ D {\bf 74}, 023510 (2006)
  doi:10.1103/PhysRevD.74.023510
  [hep-ph/0602230].

\bibitem{TuckerSmith:2001hy} 
  D.~Tucker-Smith and N.~Weiner,
  Phys.\ Rev.\ D {\bf 64}, 043502 (2001)
  doi:10.1103/PhysRevD.64.043502
  [hep-ph/0101138].

\bibitem{Faraggi:2000pv} 
  A.~E.~Faraggi and M.~Pospelov,
  Astropart.\ Phys.\  {\bf 16}, 451 (2002)
  doi:10.1016/S0927-6505(01)00121-9
  [hep-ph/0008223].

\bibitem{Juknevich:2009ji} 
  J.~E.~Juknevich, D.~Melnikov and M.~J.~Strassler,
  JHEP {\bf 0907}, 055 (2009)
  doi:10.1088/1126-6708/2009/07/055
  [arXiv:0903.0883 [hep-ph]].

\bibitem{Boddy:2014yra} 
  K.~K.~Boddy, J.~L.~Feng, M.~Kaplinghat and T.~M.~P.~Tait,
  Phys.\ Rev.\ D {\bf 89}, no. 11, 115017 (2014)
  doi:10.1103/PhysRevD.89.115017
  [arXiv:1402.3629 [hep-ph]].

\bibitem{Boddy:2014qxa} 
  K.~K.~Boddy, J.~L.~Feng, M.~Kaplinghat, Y.~Shadmi and T.~M.~P.~Tait,
  Phys.\ Rev.\ D {\bf 90}, no. 9, 095016 (2014)
  doi:10.1103/PhysRevD.90.095016
  [arXiv:1408.6532 [hep-ph]].

\bibitem{Morningstar:1999rf} 
  C.~J.~Morningstar and M.~J.~Peardon,
  Phys.\ Rev.\ D {\bf 60}, 034509 (1999)
  doi:10.1103/PhysRevD.60.034509
  [hep-lat/9901004].

\bibitem{Chen:2005mg} 
  Y.~Chen {\it et al.},
  Phys.\ Rev.\ D {\bf 73}, 014516 (2006)
  doi:10.1103/PhysRevD.73.014516
  [hep-lat/0510074].

\bibitem{Meyer:2008tr} 
  H.~B.~Meyer,
  JHEP {\bf 0901}, 071 (2009)
  doi:10.1088/1126-6708/2009/01/071
  [arXiv:0808.3151 [hep-lat]].

\bibitem{Craig:2015pha} 
  N.~Craig, A.~Katz, M.~Strassler and R.~Sundrum,
  JHEP {\bf 1507}, 105 (2015)
  doi:10.1007/JHEP07(2015)105
  [arXiv:1501.05310 [hep-ph]].

\bibitem{Curtin:2015fna} 
  D.~Curtin and C.~B.~Verhaaren,
  JHEP {\bf 1512}, 072 (2015)
  doi:10.1007/JHEP12(2015)072
  [arXiv:1506.06141 [hep-ph]].

\bibitem{Chou:2016lxi} 
  J.~P.~Chou, D.~Curtin and H.~J.~Lubatti,
  Phys.\ Lett.\ B {\bf 767}, 29 (2017)
  doi:10.1016/j.physletb.2017.01.043
  [arXiv:1606.06298 [hep-ph]].

\bibitem{Schwaller:2015gea} 
  P.~Schwaller, D.~Stolarski and A.~Weiler,
  JHEP {\bf 1505}, 059 (2015)
  doi:10.1007/JHEP05(2015)059
  [arXiv:1502.05409 [hep-ph]].

\bibitem{Cohen:2015toa} 
  T.~Cohen, M.~Lisanti and H.~K.~Lou,
  Phys.\ Rev.\ Lett.\  {\bf 115}, no. 17, 171804 (2015)
  doi:10.1103/PhysRevLett.115.171804
  [arXiv:1503.00009 [hep-ph]].

\bibitem{Baumgart:2009tn} 
  M.~Baumgart, C.~Cheung, J.~T.~Ruderman, L.~T.~Wang and I.~Yavin,
  JHEP {\bf 0904}, 014 (2009)
  doi:10.1088/1126-6708/2009/04/014
  [arXiv:0901.0283 [hep-ph]].

\bibitem{Choquette:2015mca} 
  J.~Choquette and J.~M.~Cline,
  Phys.\ Rev.\ D {\bf 92}, no. 11, 115011 (2015)
  doi:10.1103/PhysRevD.92.115011
  [arXiv:1509.05764 [hep-ph]].

\bibitem{Barello:2015bhq} 
  G.~Barello, S.~Chang and C.~A.~Newby,
  Phys.\ Rev.\ D {\bf 94}, no. 5, 055018 (2016)
  doi:10.1103/PhysRevD.94.055018
  [arXiv:1511.02865 [hep-ph]].

\bibitem{Okun:1980kw} 
  L.~B.~Okun,
  JETP Lett.\  {\bf 31}, 144 (1980)
  [Pisma Zh.\ Eksp.\ Teor.\ Fiz.\  {\bf 31}, 156 (1979)].

\bibitem{Okun:1980mu} 
  L.~B.~Okun,
  Nucl.\ Phys.\ B {\bf 173}, 1 (1980).
  doi:10.1016/0550-3213(80)90439-3

\bibitem{Kang:2008ea} 
  J.~Kang and M.~A.~Luty,
  JHEP {\bf 0911}, 065 (2009)
  doi:10.1088/1126-6708/2009/11/065
  [arXiv:0805.4642 [hep-ph]].

\bibitem{Martin:2010kk} 
  S.~P.~Martin,
  Phys.\ Rev.\ D {\bf 83}, 035019 (2011)
  doi:10.1103/PhysRevD.83.035019
  [arXiv:1012.2072 [hep-ph]].

\bibitem{Harnik:2011mv} 
  R.~Harnik, G.~D.~Kribs and A.~Martin,
  Phys.\ Rev.\ D {\bf 84}, 035029 (2011)
  doi:10.1103/PhysRevD.84.035029
  [arXiv:1106.2569 [hep-ph]].

\bibitem{Cheung:2008ke} 
  K.~Cheung, W.~Y.~Keung and T.~C.~Yuan,
  Nucl.\ Phys.\ B {\bf 811}, 274 (2009)
  doi:10.1016/j.nuclphysb.2008.11.029
  [arXiv:0810.1524 [hep-ph]].

\bibitem{Barger:1987xg} 
  V.~D.~Barger, E.~W.~N.~Glover, K.~Hikasa, W.~Y.~Keung, M.~G.~Olsson, C.~J.~Suchyta, III and X.~R.~Tata,
  Phys.\ Rev.\ D {\bf 35}, 3366 (1987)
  Erratum: [Phys.\ Rev.\ D {\bf 38}, 1632 (1988)].
  doi:10.1103/PhysRevD.35.3366, 10.1103/PhysRevD.38.1632.2

\bibitem{Chacko:2015fbc} 
  Z.~Chacko, D.~Curtin and C.~B.~Verhaaren,
  Phys.\ Rev.\ D {\bf 94}, no. 1, 011504 (2016)
  doi:10.1103/PhysRevD.94.011504
  [arXiv:1512.05782 [hep-ph]].

\bibitem{Jacoby:2007nw} 
  C.~Jacoby and S.~Nussinov,
  arXiv:0712.2681 [hep-ph].

\bibitem{Nussinov:2009hc} 
  S.~Nussinov and C.~Jacoby,
  arXiv:0907.4932 [hep-ph].

\bibitem{Kawasaki:2004qu} 
  M.~Kawasaki, K.~Kohri and T.~Moroi,
  Phys.\ Rev.\ D {\bf 71}, 083502 (2005)
  doi:10.1103/PhysRevD.71.083502
  [astro-ph/0408426].

\bibitem{Jedamzik:2006xz} 
  K.~Jedamzik,
  Phys.\ Rev.\ D {\bf 74}, 103509 (2006)
  doi:10.1103/PhysRevD.74.103509
  [hep-ph/0604251].

\bibitem{Chen:2003gz} 
  X.~L.~Chen and M.~Kamionkowski,
  Phys.\ Rev.\ D {\bf 70}, 043502 (2004)
  doi:10.1103/PhysRevD.70.043502
  [astro-ph/0310473].

\bibitem{Fradette:2014sza} 
  A.~Fradette, M.~Pospelov, J.~Pradler and A.~Ritz,
  Phys.\ Rev.\ D {\bf 90}, no. 3, 035022 (2014)
  doi:10.1103/PhysRevD.90.035022
  [arXiv:1407.0993 [hep-ph]].

\bibitem{Garcia:2015loa} 
  I.~Garcia Garcia, R.~Lasenby and J.~March-Russell,
  Phys.\ Rev.\ D {\bf 92}, no. 5, 055034 (2015)
  doi:10.1103/PhysRevD.92.055034
  [arXiv:1505.07109 [hep-ph]].

\bibitem{Forestell:2016qhc} 
  L.~Forestell, D.~E.~Morrissey and K.~Sigurdson,
  Phys.\ Rev.\ D {\bf 95}, no. 1, 015032 (2017)
  doi:10.1103/PhysRevD.95.015032
  [arXiv:1605.08048 [hep-ph]].

\bibitem{Soni:2017nlm} 
  A.~Soni, H.~Xiao and Y.~Zhang,
  arXiv:1704.02347 [hep-ph].

\bibitem{Voloshin:2012tv} 
  M.~B.~Voloshin,
  Phys.\ Rev.\ D {\bf 86}, 093016 (2012)
  doi:10.1103/PhysRevD.86.093016
  [arXiv:1208.4303 [hep-ph]].

\bibitem{Passarino:1978jh} 
  G.~Passarino and M.~J.~G.~Veltman,
  Nucl.\ Phys.\ B {\bf 160}, 151 (1979).
  doi:10.1016/0550-3213(79)90234-7

\end{thebibliography}

\end{document}